\documentclass[%
reprint,
superscriptaddress,
 amsmath,amssymb,
 aps,
]{revtex4-2}

\usepackage{graphicx}
\usepackage{dcolumn}
\usepackage{bm}

\usepackage[english]{babel}
\usepackage[caption=false]{subfig}
\usepackage{siunitx, soul, xcolor}
\usepackage{adjustbox}
\usepackage{gensymb}

\usepackage{xparse,xcoffins}

\ExplSyntaxOn
\NewCoffin\imagecoffin
\NewCoffin\labelcoffin

\keys_define:nn { miguel/label }
 {
  label   .tl_set:N = \l_miguel_label_tl,
  labelbox .bool_set:N = \l_miguel_label_box_bool,
  labelbox .default:n = true,
  fontsize .tl_set:N = \l_miguel_label_size_tl,
  fontsize .initial:n = \footnotesize,
  pos .choice:,
  pos/nw .code:n = \tl_set:Nn \l_miguel_label_pos_tl { left,up },
  pos/ne .code:n = \tl_set:Nn \l_miguel_label_pos_tl { right,up },
  pos/sw .code:n = \tl_set:Nn \l_miguel_label_pos_tl { left,down },
  pos/se .code:n = \tl_set:Nn \l_miguel_label_pos_tl { right,down },
  pos/n .code:n = \tl_set:Nn \l_miguel_label_pos_tl { hc,up },
  pos/w .code:n = \tl_set:Nn \l_miguel_label_pos_tl { left,vc },
  pos/s .code:n = \tl_set:Nn \l_miguel_label_pos_tl { hc,down },
  pos/e .code:n = \tl_set:Nn \l_miguel_label_pos_tl { right,vc },
  pos .initial:n = nw,
  unknown .code:n   = \clist_put_right:Nx \l_miguel_label_clist
                       { \l_keys_key_tl = \exp_not:n { #1 } }
 }
\clist_new:N \l_miguel_label_clist
\box_new:N \l_miguel_label_box
\box_new:N \l_miguel_label_image_box

\NewDocumentCommand{\xincludegraphics}{O{}m}
 {
  \group_begin:
  \tl_clear:N \l_miguel_label_tl
  \clist_clear:N \l_miguel_label_clist
  \keys_set:nn { miguel/label } { #1 }
  \tl_if_empty:NTF \l_miguel_label_tl
   {
    \miguel_includegraphics:Vn \l_miguel_label_clist { #2 }
   }
   {
    \SetHorizontalCoffin\imagecoffin
     {
      \miguel_includegraphics:Vn \l_miguel_label_clist { #2 }
     }
    \SetHorizontalCoffin\labelcoffin
     {
      \raisebox{\depth}
       {
        \bool_if:NTF \l_miguel_label_box_bool
         { \fcolorbox{white}{white}{\l_miguel_label_size_tl\l_miguel_label_tl} }
         { \l_miguel_label_size_tl\l_miguel_label_tl }
       }
     }
    \SetVerticalPole\imagecoffin{left}{-2pt+\CoffinWidth\labelcoffin/2}
    \SetVerticalPole\imagecoffin{right}{\Width-3pt-\CoffinWidth\labelcoffin/2}
    \SetHorizontalPole\imagecoffin{up}{\Height+5pt-\CoffinHeight\labelcoffin/2}
    \SetHorizontalPole\imagecoffin{down}{3pt+\CoffinHeight\labelcoffin/2}
    \use:x{\JoinCoffins\imagecoffin[\l_miguel_label_pos_tl]\labelcoffin[vc,hc]} 
    \TypesetCoffin\imagecoffin
   }
   \group_end:
 }
\NewDocumentCommand{\setlabel}{m}
 {
  \keys_set:nn { miguel/label } { #1 }
 }

\cs_new_protected:Nn \miguel_includegraphics:nn
 {
  \includegraphics[#1]{#2}
 }
\cs_generate_variant:Nn \miguel_includegraphics:nn { V }

\ExplSyntaxOff

\begin{document}
\preprint{APS/123-QED}

\title{Ultrafast Light-Induced Magnetoelectric Effect in van der Waals \\ Magnetic Semiconductor Heterostructures}

\author{Wenyi Zhou}
\affiliation{The Ohio State University, Department of Physics, Columbus, OH, USA}
\author{Ravi Kumar Bandapelli}
\affiliation{Carnegie Mellon University, Department of Physics, Pittsburgh, PA, USA}
\author{Hari Paudyal}
\affiliation{University of Iowa, Department of Physics and Astronomy, Iowa City, IA, USA}
\author{Bangzheng Han}
\affiliation{The Ohio State University, Department of Physics, Columbus, OH, USA}
\author{I-Hsuan Kao}
\affiliation{Carnegie Mellon University, Department of Physics, Pittsburgh, PA, USA}
\author{Ziling Li}
\affiliation{The Ohio State University, Department of Physics, Columbus, OH, USA}
\author{Yuqing Zhu}
\affiliation{The Ohio State University, Department of Physics, Columbus, OH, USA}
\author{Durga Paudyal}
\affiliation{University of Iowa, Department of Physics and Astronomy, Iowa City, IA, USA}
\author{Jyoti Katoch}
\affiliation{Carnegie Mellon University, Department of Physics, Pittsburgh, PA, USA}
\author{Simranjeet Singh}
\affiliation{Carnegie Mellon University, Department of Physics, Pittsburgh, PA, USA}
\author{Roland K. Kawakami}
\email[email address: ]{kawakami.15@osu.edu}
\affiliation{The Ohio State University, Department of Physics, Columbus, OH, USA}

\begin{abstract}
Atomic-scale heterostructures of van der Waals (vdW) magnets and semiconductors provide a unique environment for exploring magnetic dynamics.
In contrast to typical photothermal excitation of precessional magnetization dynamics by a pump laser pulse, we find that ultrafast optical excitation of a WS$_2$/CrGeTe$_3$ (CGT) bilayer produces an opposite sign of magnetic torque compared to an isolated CGT film. Experimental observations by time-resolved magneto-optic Kerr effect (TR-MOKE) and theoretical analysis by density functional theory (DFT) and Landau-Lifshitz-Gilbert (LLG) simulations support a mechanism in which charge transfer of photoexcited carriers across the interface alters the perpendicular magnetic anisotropy, which in turn generates a torque on the magnetic layer to trigger precessional magnetization dynamics. 
These results provide new avenues for ultrafast manipulation of magnetization in vdW heterostructures with type-II band alignments. Lastly, we show that optically-generated spin currents from WS$_2$ into CGT can also trigger precessional dynamics via angular momentum transfer.

\end{abstract}

\keywords{2D magnets, van der Waals heterostructure, magnetization dynamics, time-resolved Kerr ellipticity, magnetic anisotropy}

\maketitle

The use of ultrafast optical pulses for triggering magnetic dynamics has generated substantial interest for rapid all-optical magnetization switching \cite{kirilyuk_ultrafast_2010, dabrowski_all-optical_2022}, launching coherent spin waves \cite{zhang_gate-tunable_2020, bae_exciton-coupled_2022, diederich_tunable_2023, sun_dipolar_2024, bartram_real-time_2023}, and generating terahertz radiation \cite{rongione_emission_2023, li_above-curie-temperature_2025}.
These could be important for developing applications in fast magnetic memories and energy-efficient devices for magnonics \cite{kimel_writing_2019, barman_2021_2021}, but a challenge is to enhance the coupling between the photonic and magnetic degrees of freedom.

A typical mechanism for triggering dynamics of a ferromagnetic thin film involves thermal excitation by absorption of a laser pulse, which modifies the effective magnetic field through ultrafast demagnetization and/or thermally-induced changes to the magnetocrystalline anisotropy \cite{beaurepaire_ultrafast_1996, iihama_gilbert_2014, tang_spin_2023}. As a result, the total magnetic field becomes momentarily misaligned with the magnetization vector, leading to the excitation of precessional dynamics \cite{zhang_laser-induced_2020, hendriks_electric_2024}.
Another mechanism is a non-thermal effect where the helicity of the pump pulse generates precessional dynamics through the transfer of angular momentum from the photons to the magnetic moments \cite{kimel_ultrafast_2005, choi_optical-helicity-driven_2017}.

The advent of atomic-scale van der Waals (vdW) heterostructures provides new opportunities to enhance the optical triggering of magnetization dynamics \cite{zhang_gate-tunable_2020}. 
Monolayer transition metal dichalcogenides (TMDs) such as WS$_2$ are direct gap semiconductors with large exciton binding and strong optical absorption peaks \cite{chernikov_exciton_2014, mak_photonics_2016,  aivazian_magnetic_2015, zhong_van_2017, norden_giant_2019, seyler_valley_2018, lyons2020interplay, choi_asymmetric_2023}. 
vdW magnets can remain magnetic down to single vdW layers, which lowers the magnetic volume for enhanced sensitivity to interfacial interactions \cite{huang_layer-dependent_2017, gong_discovery_2017, zhuo_manipulating_2021, tu_spinorbit_2022, gupta_manipulation_2020, goff_scanning_2024, cham_spin-filter_2025}. In addition, the inherent structural anisotropy (in-plane vs.~out-of-plane) naturally permits the generation of perpendicular magnetic anisotropy \cite{gibertini_magnetic_2019}.
Recent experiments using the time-resolved magneto-optic Kerr effect (TR-MOKE) on CrGeTe$_3$ (CGT), a ferromagnetic semiconductor \cite{gong_discovery_2017, zhang_laser-induced_2020, sun_ultra-long_2021, sutcliffe_transient_2023, dabrowski_ultrafast_2025}, showed that precessional dynamics could be excited thermally by an optical pump pulse, and the precession frequency was controlled using electrostatic gates \cite{hendriks_electric_2024}.

In this Letter, we find that the optical triggering of magnetization dynamics in WS$_2$/CGT bilayers is greatly enhanced compared to the excitation of CGT alone. Using TR-MOKE, we observe precessional dynamics with a much larger amplitude for WS$_2$/CGT and a $\sim$180$\degree$ relative phase shift between WS$_2$/CGT and CGT, indicating that the optically-induced magnetic torque has opposite signs in the two cases.
These observations suggest an alternative mechanism for exciting magnetization precession, and further experiments and theoretical analysis support the presence of an ultrafast light-induced magnetoelectric effect that triggers the dynamics in WS${_2}$/CGT.
In this mechanism, the optical excitation produces rapid charge transfer between WS$_2$ and CGT, which modifies the carrier density in the layers and generates an electric field across the interface. These modifications of the charge distribution and electric field induce a change in the magnetic anisotropy, which is viewed as a light-induced magnetoelectric effect. These variations in the magnetic anisotropy are consistent with observations from previous electrostatic-gating studies of CGT \cite{verzhbitskiy_controlling_2020,hendriks_electric_2024} and are supported by density functional theory (DFT) calculations of the magnetic anisotropy. The ultrafast change in the magnetic anisotropy, in turn, modifies the total effective magnetic field and triggers the magnetization dynamics.
The overall mechanism relies on type-II band alignment between WS$_2$ and CGT, which is supported by DFT calculations, photocurrent measurements, and pump-wavelength dependence of the TR-MOKE experiments.
Lastly, we address an important open question of whether optically-generated spin polarization in the TMD layer transfers its angular momentum to an adjacent magnetic layer. 
By modulating the pump helicity, we demonstrate that ultrafast optical generation of valley and spin polarization in WS$_{2}$ triggers magnetization dynamics in CGT through transfer of angular momentum.

\begin{figure}[h]
\setlabel{pos=nw,fontsize=\scriptsize} 

  \subfloat{\xincludegraphics[width=0.18\textwidth,label=\textcolor{black}{(a)}]{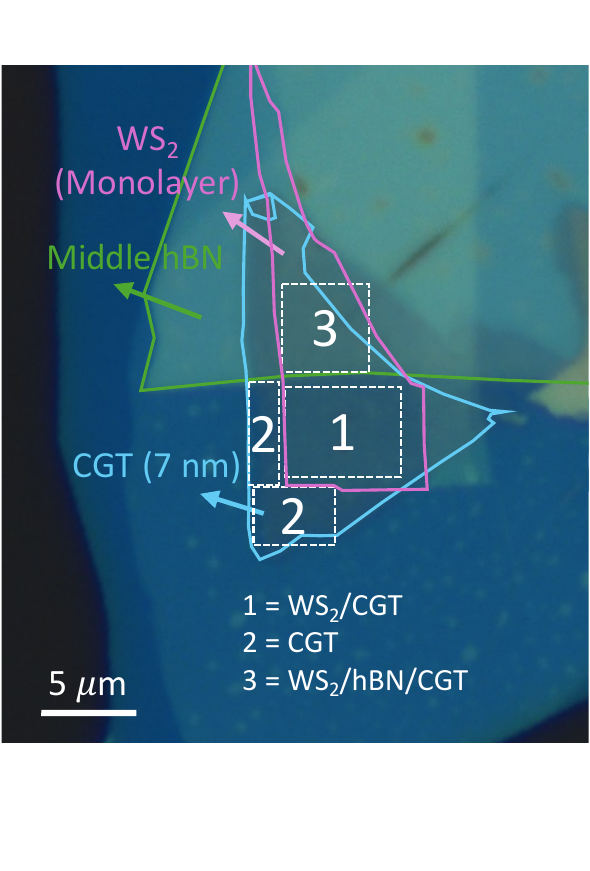}
  \label{fig:optical_stack}}
  \subfloat{\xincludegraphics[width=0.285\textwidth,label=\textcolor{black}{(b)}]{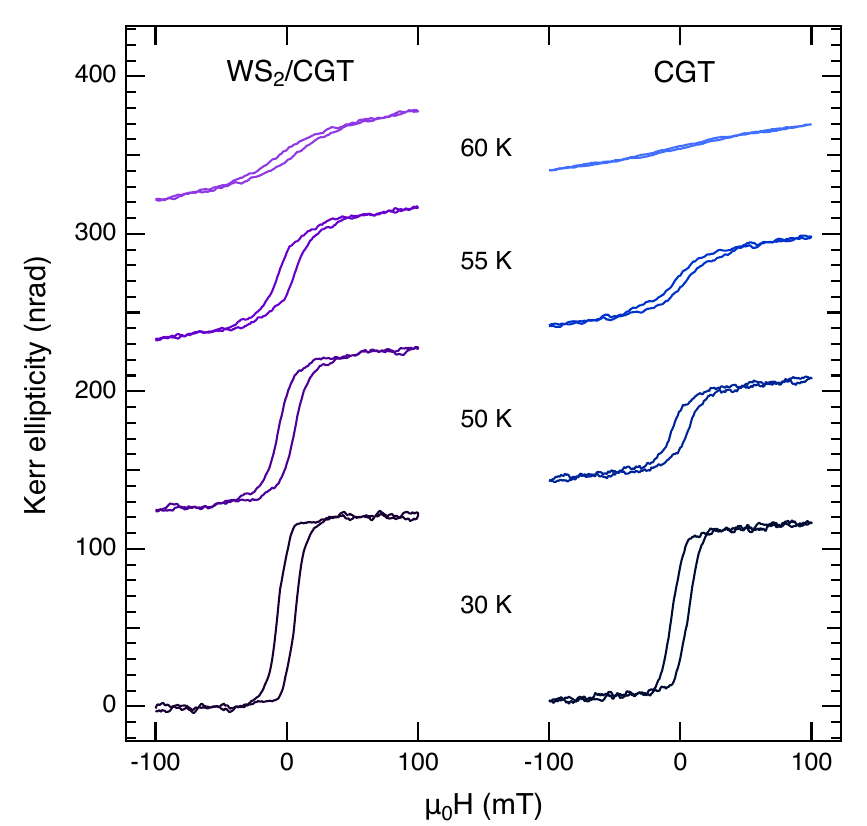}
  \label{fig:hysteresis_loop}}
  \hfill
  \par\vspace{-0.2cm} 
  \subfloat{\xincludegraphics[width=0.15\textwidth,label=\textcolor{black}{(c)}]{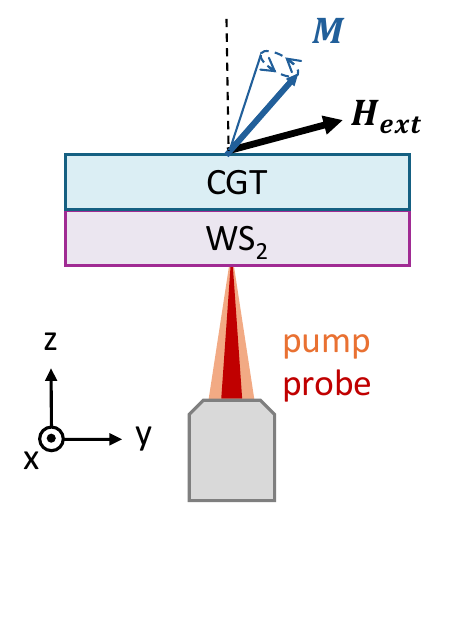}\label{fig:schematic_precession_pump_probe}}
  \hfill
  \subfloat{\xincludegraphics[width=0.3\textwidth,label=\textcolor{black}{(d)}]{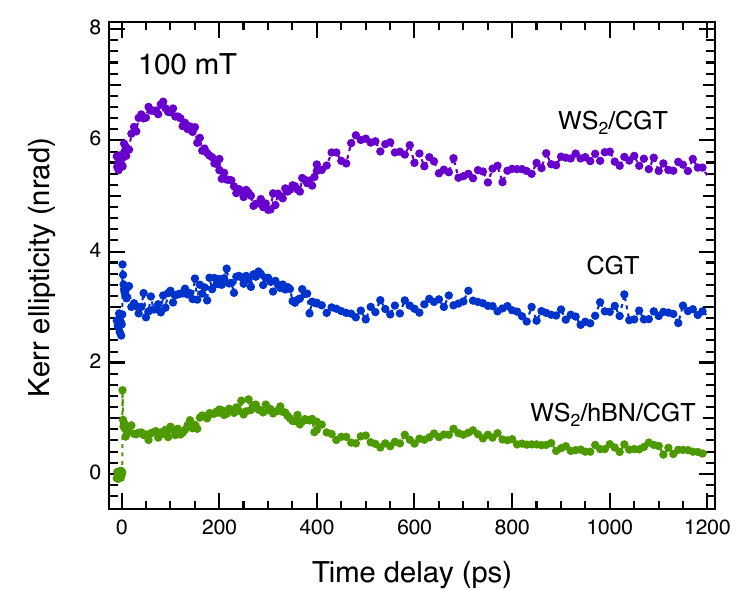}\label{fig:compare_precession_40K}}
  \hfill
  
\caption{\textbf{Sample geometry and characterization.} 
(a) Optical microscope image of the stacked heterostructure showing three distinct regions, (1) WS$_2$/Cr$_2$Ge$_2$Te$_6$ (CGT), (2) CGT, (3) WS$_2$/hexagonal boron nitride (hBN)/CGT. The sample is encapsulated with an hBN top layer and hBN bottom layer. (b) Out-of-plane hysteresis loops of the WS$_2$/CGT and CGT measured by the Kerr ellipticity (KE). 
(c) Schematic of pump-probe measurement on the heterostructure WS$_2$/CGT. (d) Typical TRKE delay scans on WS$_2$/CGT (purple), CGT (blue) and WS$_2$/hBN/CGT (green) at 40\,K with external field of 100\,mT.
}\label{fig:Fig1sketch}
\end{figure}

Fig.~\ref{fig:optical_stack} shows an optical microscopy image of the WS$_2$/CGT heterostructure sample. The heterostructure was fabricated by stacking a monolayer WS$_2$ on a CGT flake on SiO$_2$/Si substrate. The whole sample was encapsulated by top and bottom hBN. The sample contains several co-existing regions: (1) WS$_2$/CGT; (2) CGT; and (3) WS$_2$/hBN/CGT, enabling a comparison of different configurations for the experiments.

To confirm and compare ferromagnetic order in CGT and the WS$_2$/CGT heterostructure, we perform static Kerr ellipticity measurements on the different regions. Out-of-plane magnetization is probed using normal-incidence geometry, where the polar Kerr ellipticity (KE) is recorded as a function of applied out-of-plane magnetic field. As shown in Fig.\ref{fig:hysteresis_loop}, both regions exhibit clear hysteresis at $\sim$30\,K, confirming ferromagnetic ordering with perpendicular magnetic anisotropy (PMA).
The hysteresis loops of WS$_2$/CGT heterostructure are more square and the ferromagnetism persists up to higher temperatures compared to CGT alone. From detailed temperature dependent measurements (see Supplemental Material (SM)~\cite{SupplMat}), the Curie temperature of the WS$_2$/CGT bilayer is determined to be $\sim$65\,K, slightly higher than the $\sim$61\,K observed for CGT alone. These observations indicate slightly enhanced PMA in the WS$_2$/CGT heterostructure.

Fig.~\ref{fig:schematic_precession_pump_probe} shows a schematic of the TRKE measurements on WS$_2$/CGT.
Precessional dynamics in CGT is triggered by a 598\,nm pump pulse (resonant with WS$_2$) and is detected by a time delayed 800\,nm probe pulse via measuring the z-component of magnetization. Measurement details are described in SM. 
In Fig.~\ref{fig:compare_precession_40K}, we compare the typical TRKE delay scans for WS$_2$/CGT (purple curve) and CGT (blue curve). The precessional frequencies are very similar, while the most prominent feature is the large difference in starting phase. This suggests possible different mechanisms for exciting the magnetization dynamics. A second prominent difference is that
the precession amplitude is significantly higher when the CGT is interfaced with WS$_2$.
We also investigate the magnetization dynamics in WS$_2$/hBN/CGT (green curve in Fig.~\ref{fig:compare_precession_40K}) where the insulating hBN blocks charge transfer and eliminates any interfacial proximity effects. The similarity of dynamics with CGT (blue) and the difference with WS$_2$/CGT (purple) suggest the importance of charge transfer or proximity effects in the WS$_2$/CGT case.

\begin{figure}[!t]
\setlabel{pos=nw,fontsize=\scriptsize} 
\raggedright \subfloat{\xincludegraphics[width=0.22\textwidth,label=\textcolor{black}{(a)}]{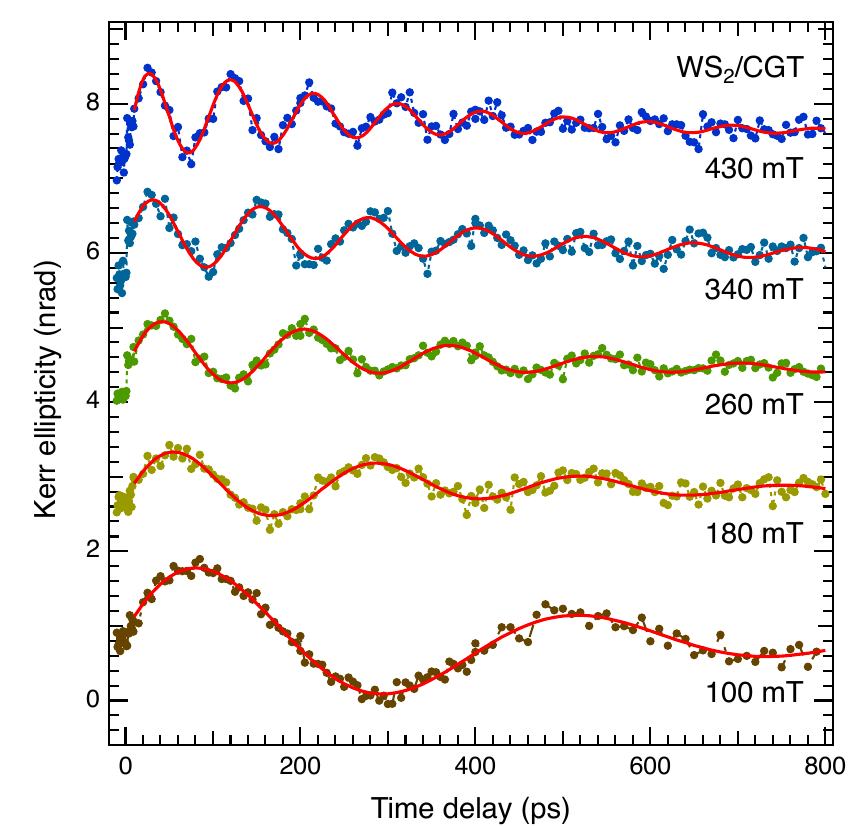}\label{fig:WS2_CGT_40K_598nm}}
\subfloat{\xincludegraphics[width=0.22\textwidth,label=\textcolor{black}{(b)}]{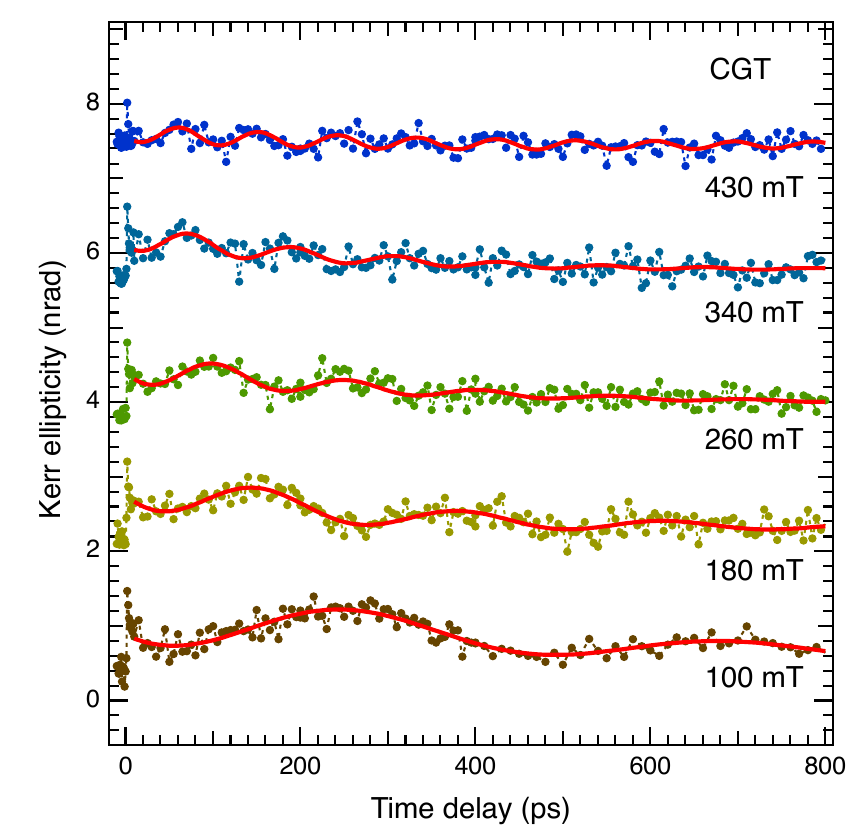}\label{fig:CGT_40K_598nm}}
\par\vspace{-0.3cm} 
\subfloat{\xincludegraphics[width=0.44\textwidth,label=\textcolor{black}{(c)}]{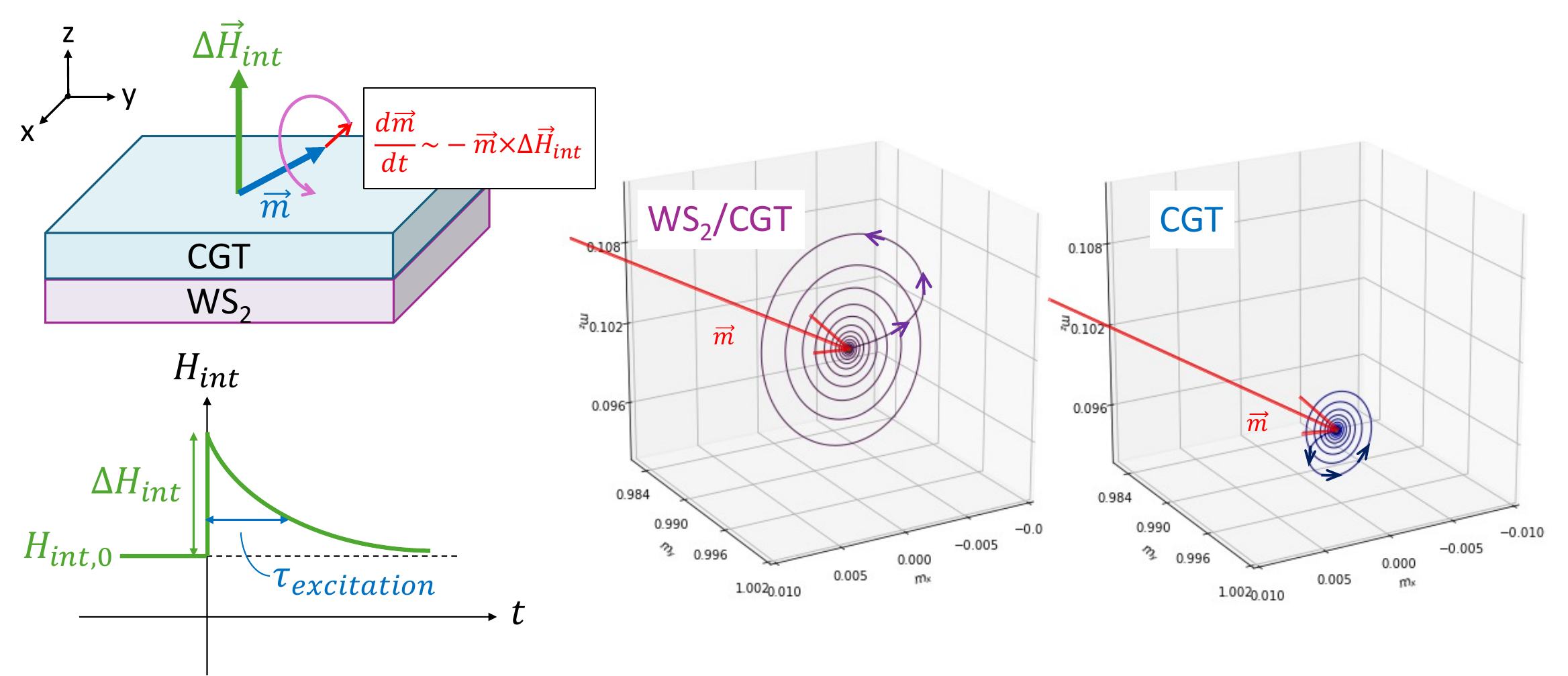}\label{fig:LLG_all}
}
\par\vspace{-0.2cm} 
\subfloat{\xincludegraphics[width=0.2\textwidth,label=\textcolor{black}{(d)}]{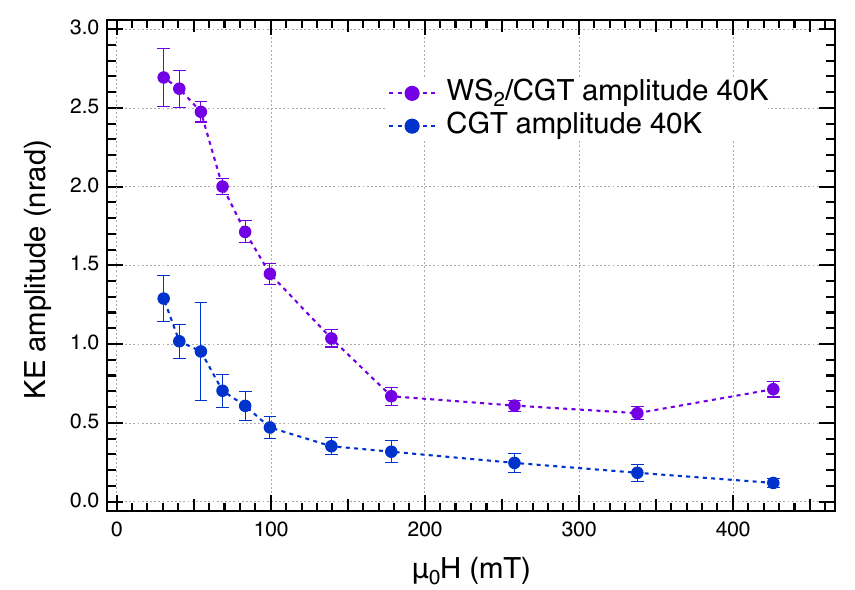}\label{fig:a_vs_H_40K}
}
\subfloat{\xincludegraphics[width=0.2\textwidth,label=\textcolor{black}{(e)}]{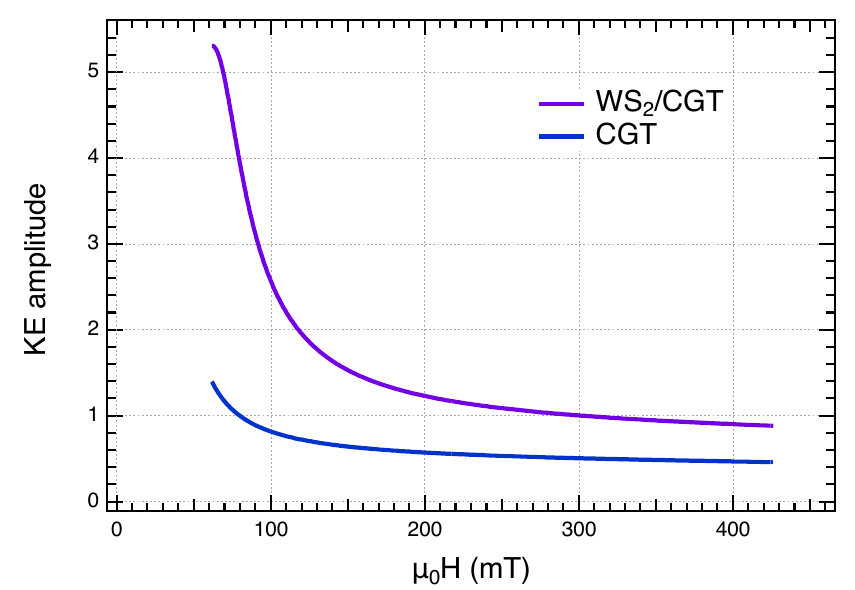}\label{fig:a_vs_H_LLG}
}
\par\vspace{-0.4cm} 

\subfloat{\xincludegraphics[width=0.2\textwidth,label=\textcolor{black}{(f)}]{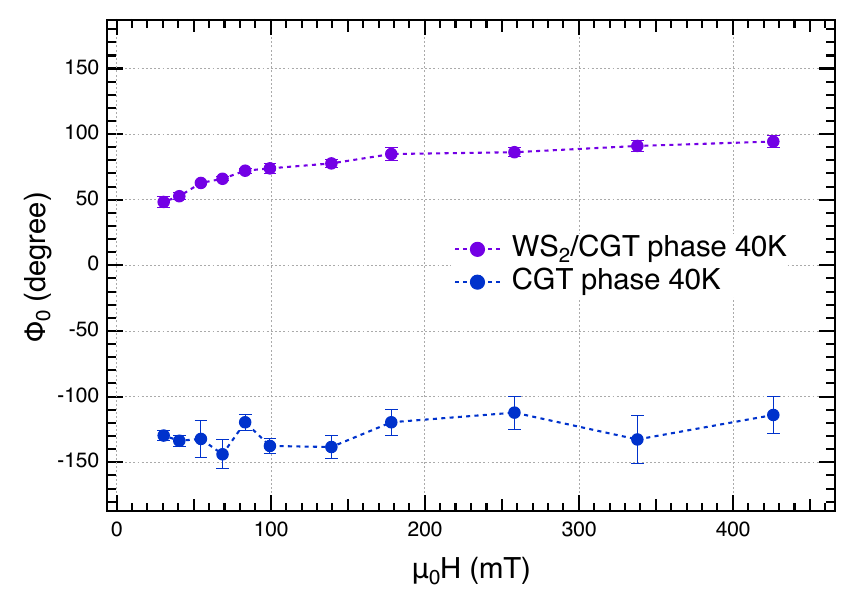}\label{fig:phi_vs_H_40K}
}
\subfloat{\xincludegraphics[width=0.2\textwidth,label=\textcolor{black}{(g)}]{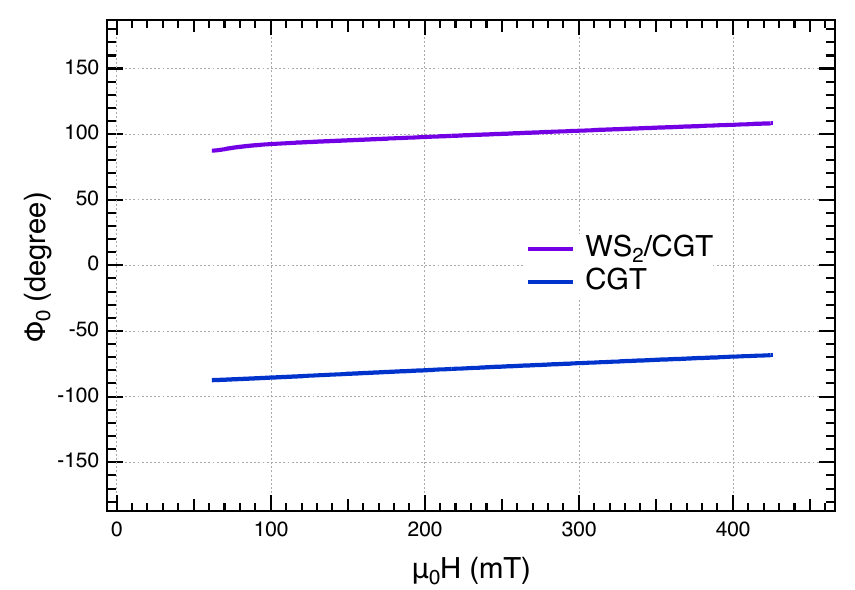}\label{fig:phi_vs_H_LLG}
}
\par\vspace{-0.4cm} 

\subfloat{\xincludegraphics[width=0.2\textwidth,label=\textcolor{black}{(h)}]{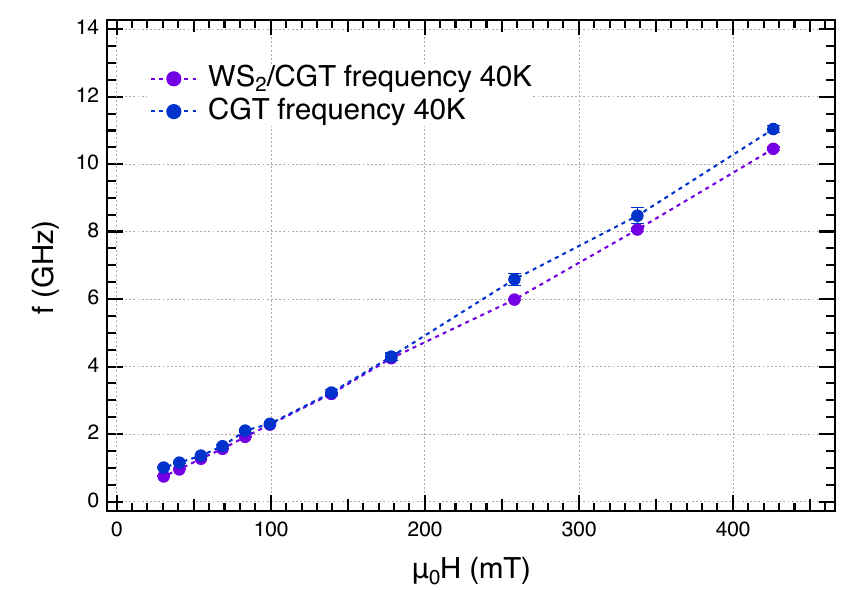}\label{fig:f_vs_H_40K}
}
\subfloat{\xincludegraphics[width=0.2\textwidth,label=\textcolor{black}{(i)}]{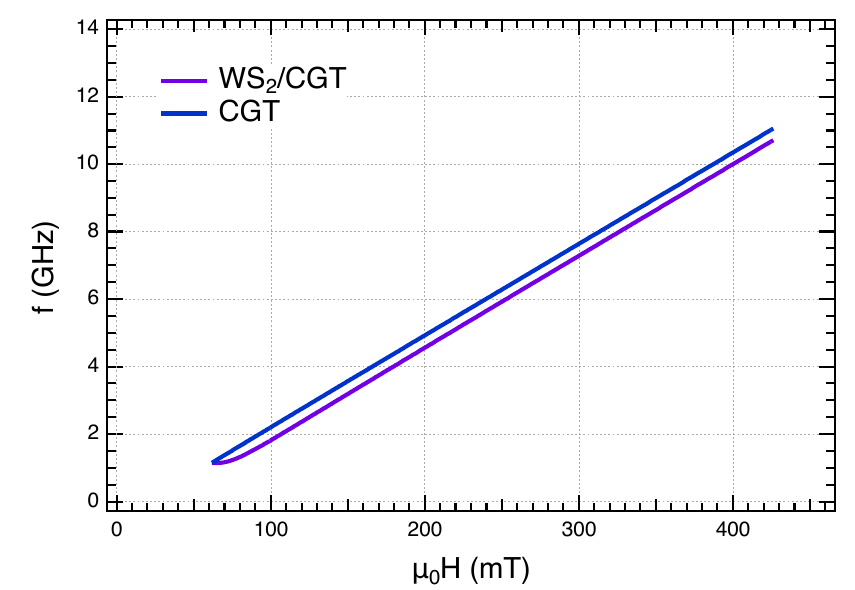}\label{fig:f_vs_H_LLG}
}
\caption{\textbf{Time-resolved magnetization dynamics and LLG equation modelling.} (a,b) TRKE delay scans of WS$_2$/CGT and CGT respectively, at various magnetic fields. The red solid lines represent fitting curves.
(c) Illustration of LLG equation modelling with pump-induced $\Delta H_{int}$ (left) and simulated magnetization trajectory for WS$_2$/CGT (middle) and CGT (right) with optimized parameters.
(d,f,h) Extracted precession amplitude $a$, phase $\phi_0$ and frequency $f$ as a function of external field $H_{ext}$ for the two regions. (e,g,i) Simulated $a$, $\phi_0$ and $f$ as a function of external field $H_{ext}$ using LLG equation modelling.
}
\label{fig:Fig2sketch}
\end{figure}

Fig.~\ref{fig:WS2_CGT_40K_598nm} and \ref{fig:CGT_40K_598nm} displays the evolution of the pump-induced magnetization precessional dynamics detected by TRKE with different external fields $\mu_0H_{ext}$ in the range of 100\,mT to 430\,mT for the WS$_2$/CGT and CGT regions, respectively. The precession frequency gradually increases with higher H$_{ext}$. In the meantime, the precession amplitude enhancement in WS$_2$/CGT versus CGT and the substantial phase shift are consistently observed at every external field.
TRKE data are plotted together with the fitting curves (red curves) whose sinusoidal part is given by $a \,e^{-t/\tau_{s}} cos(2\pi ft - \phi_0)$,\cite{zhang_laser-induced_2020,hendriks_electric_2024}
where $a$ denotes the magnetization precession amplitude as we measured in KE, which leads to precessional cone angle of $\sim$1 deg here based on magnitude of $\sim$50 nrad in KE we get from the polar hysteresis loop (Fig.\ref{fig:hysteresis_loop}). And $\tau_{s}$, $f$, and $\phi_0$ denote spin lifetime, precession frequency, and the starting phase, respectively, and $t$ is the pump-probe time delay. Details about the background in the fitting are in the SM.

In order to understand the different magnetization dynamics observed for the WS$_2$/CGT heterostructure and for CGT alone, we use a Landau-Lifshitz-Gilbert (LLG) equation based model to simulate the excitation of magnetization dynamics (Fig.~\ref{fig:LLG_all}), together with the sinusoidal fitting above to analyze the underlying mechanism.

By extracting $a$, $\phi_0$ and $f$ at different magnetic fields, the phase difference and amplitude enhancement are observed across the entire field range (Fig.~\ref{fig:a_vs_H_40K},~\ref{fig:phi_vs_H_40K}) (see SM for more temperatures). 
Additionally, the slightly lower precession frequency observed in the WS$_2$/CGT region (Fig.~\ref{fig:f_vs_H_40K}) confirms its enhanced PMA compared to CGT alone, as we have also observed in the hysteresis loops for the two regions (Fig.~\ref{fig:hysteresis_loop}).

Now we introduce the LLG equation model used to simulate the excitation of magnetization dynamics (Fig.~\ref{fig:LLG_all})
$$\dfrac{d\mathbf{m}}{dt}=-\gamma \mu_0 \mathbf{m} \times \mathbf{H} + \alpha \mathbf{m} \times \dfrac{d\mathbf{m}}{dt}$$
where $\mathbf{m}$ is the magnetization, $\gamma$ is the gyromagnetic ratio, $\alpha$ is the Gilbert damping parameter. The total magnetic field $\mathbf{H}$ is a sum of the external field $\mathbf{H_{ext}}$ and internal field $\mathbf{H_{int}} = H_{int}\cos(\theta_M)\hat{\mathbf{z}}$ with $H_{int}=-M_{eff}=\dfrac{2K_U}{\mu_0 M_S} - M_S$ arising from the perpendicular magnetocrystalline anisotropy $K_U$ and magnetic shape anisotropy. \cite{hendriks_electric_2024} Here, $M_S$ is the saturation magnetization and $\theta_M$ is the angle between $\mathbf{m}$ and sample normal ($\hat{\mathbf{z}}$). 

To model the effect of the pump excitation, we treat $H_{int}(t)$ as consisting of a time-independent part $H_{int,0}$ and a pump-induced perpendicular anisotropy field $\Delta H_{int}$ that decays exponentially: $H_{int}(t) = H_{int,0} + \Delta H_{int}\, e^{-t/\tau_{excitation}}$. 
Upon arrival of the pump pulse, $\Delta H_{int}$ dominates the precessional term of the LLG equation and produces a torque that quickly rotates the magnetization away from equilibrium (red arrow in Fig.~\ref{fig:LLG_all}):
$$\frac{d\mathbf{m}}{dt} \approx -\gamma \mu_0 \mathbf{m} \times \mathbf{\Delta H_{int}}$$
where $\mathbf{\Delta H_{int}}= \Delta H_{int} \cos\theta_{M}\,\mathbf{\hat{z}}$. As a result, opposite signs of the $\Delta H_{int}$ will lead to torques in opposite directions and produce $\sim$180$\degree$ phase shifts in the starting phase. 
As illustrated in the two trajectory plots in Fig.~\ref{fig:LLG_all}, positive values of $\Delta H_{int}$ produce $\phi_0$ near $+90\degree$ as observed for WS$_2$/CGT, while negative values of $\Delta H_{int}$ produce $\phi_0$ near $-90\degree$ as observed for CGT.
Moreover, Figures \ref{fig:a_vs_H_LLG}, \ref{fig:phi_vs_H_LLG} and \ref{fig:f_vs_H_LLG} show that  LLG simulations with appropriate choices for $H_{int,0}$, $\Delta H_{int}$ and $\tau_{excitation}$ can capture the experimentally observed trends for $a$, $\phi_{0}$ and $f$ as a function of external magnetic field (see SM for further details).
This agreement provides support that the main effect of the pump pulse is to induce an ultrafast change in the perpendicular magnetic anisotropy (i.e.~$\Delta H_{int}$$\, e^{-t/\tau_{excitation}}$).

\begin{figure}[h]
\setlabel{pos=nw,fontsize=\scriptsize} 
  \subfloat{\xincludegraphics[width=0.48\textwidth,label=\textcolor{black}{(a)}]{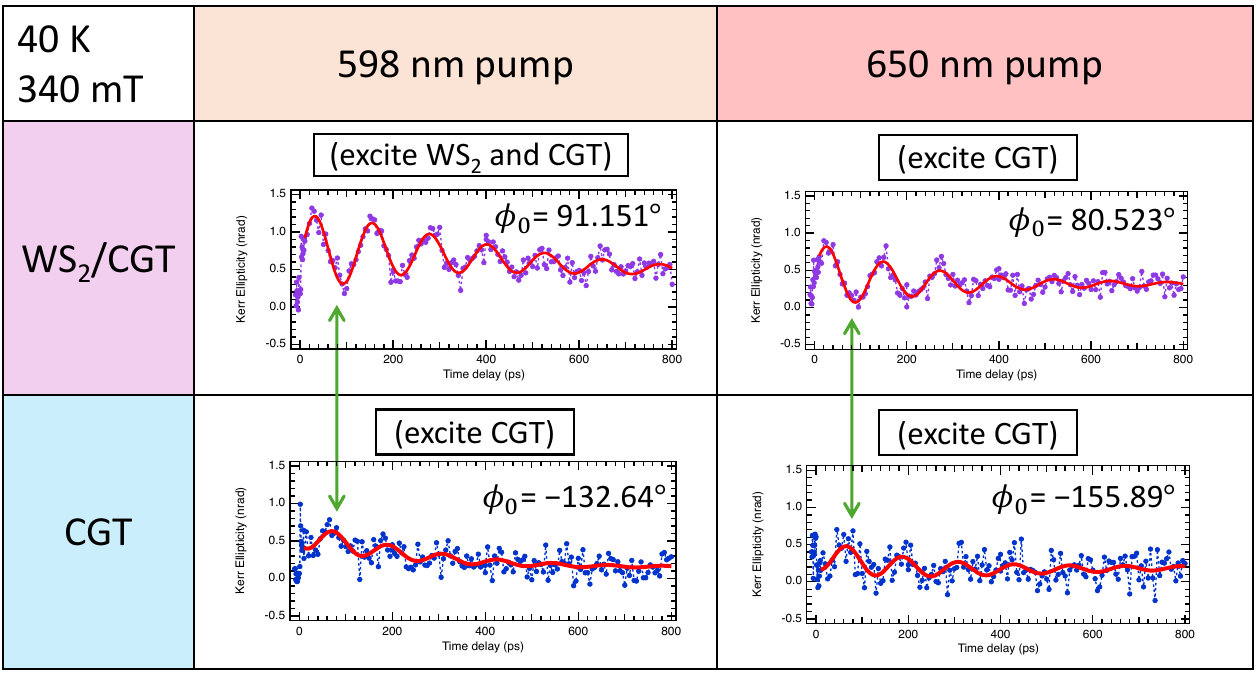}\label{fig:table}}
  \hfill
  \setlabel{pos=nw,fontsize=\scriptsize}
  \subfloat{\xincludegraphics[width=0.3\textwidth,label=\textcolor{black}{(b)}]{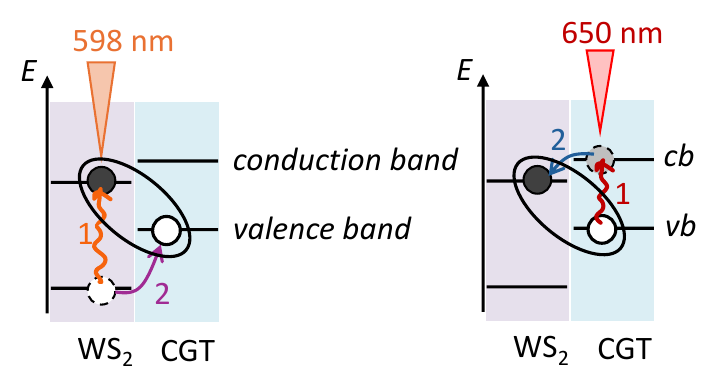}\label{fig:interlayer_exciton}}
  \vspace{-0.3cm}
  \subfloat{\xincludegraphics[width=0.45\textwidth,label=\textcolor{black}{(c)}]{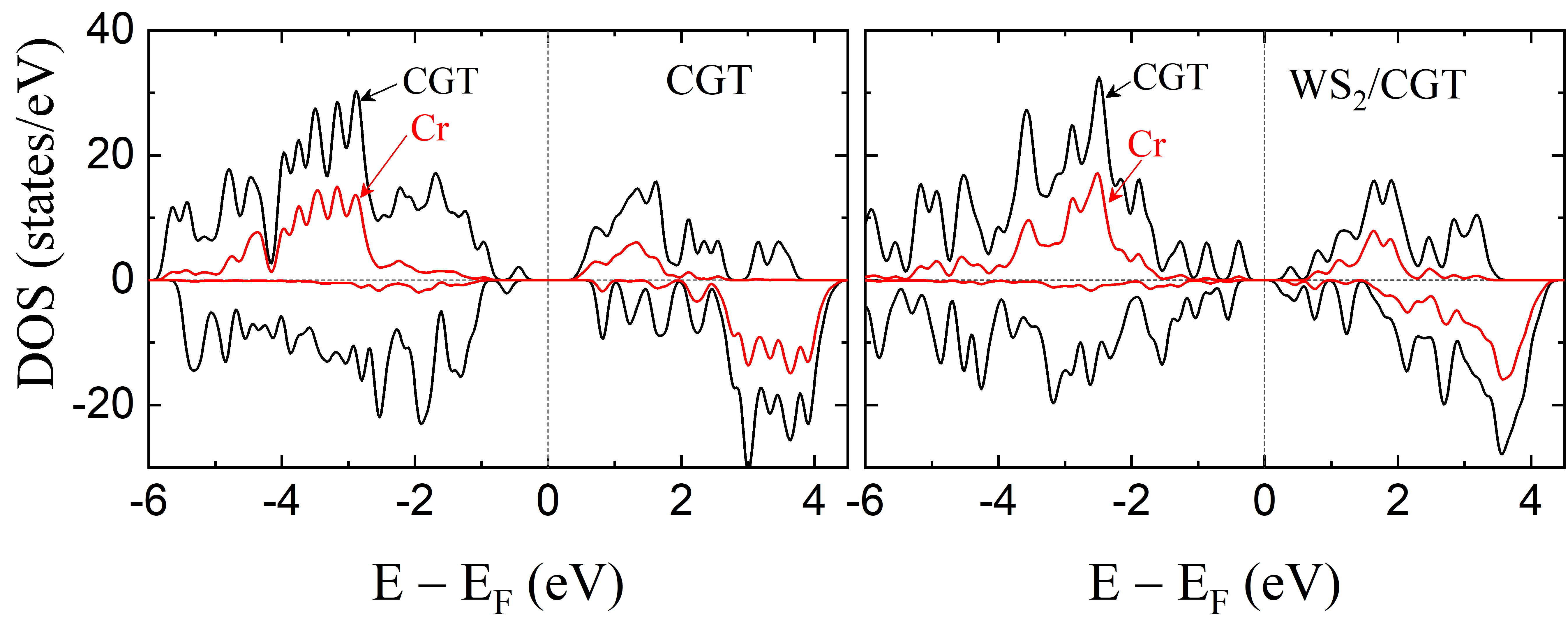}\label{fig:dftdos}}
\caption{\textbf{Pump wavelength dependence and mechanism. Density functional theory (DFT) calculations.} (a) Comparison of TRKE response with 598\,nm and 650\,nm pump on WS$_2$/CGT and CGT respectively.
(b) Schematic of type-II band alignment and pump-induced charge transfer in WS$_2$/CGT heterostructure. (c) DFT calculated projected density of states (DOS) of CGT and WS$_2$/CGT. It confirms the CGT band alignment shown in (b). Further, DFT modeling with one electron charge transfer enhances PMA by 5.2~meV in WS$_2$/CGT heterostructure.}
\label{fig:wavaelength}
\end{figure}

To investigate the origin of the pump-pulse-induced $\Delta H_{int}$$\, e^{-t/\tau_{excitation}}$, we tune the pump wavelength from 598\,nm to 650\,nm (photon energy is below the WS$_2$ absorption edge) to confirm whether the opposite starting phase and enhanced precession amplitude are related to resonant pumping in WS$_2$. In Fig.~\ref{fig:table}, we show the direct comparison of the magnetization dynamics detected by TRKE between WS$_2$/CGT and CGT using the two different pumping wavelengths under the same conditions (40\,K, 340\,mT). It turns out that the substantial phase difference between WS$_2$/CGT and CGT still exists with 650\,nm pump.
In fact, both $a$ and $\phi_0$ dependence on external field $H_{ext}$ are highly consistent with data in Fig.~\ref{fig:a_vs_H_40K} and \ref{fig:phi_vs_H_40K} where 598\,nm pump is used (see SM). Thus, it turns out that the resonant excitation in WS$_2$ is not the critical factor in determining the sign of the pump-induced torque. Instead, the presence or absence of the WS$_2$/CGT interface is the crucial factor, regardless of which layer is optically excited by the pump pulse.

A possible explanation emerges if this WS$_2$/CGT heterostructure has a type-II band alignment, as discussed in the literature \cite{rahman_giant_2021, zhang_electrically_2022} and shown in Fig.~\ref{fig:interlayer_exciton}. For both 598\,nm (exciting WS$_2$ and CGT) and 650\,nm (exciting CGT), 
the excitation is followed by rapid charge transfer of electrons from CGT to WS$_2$ and/or holes from WS$_2$ to CGT. 
This produces excess positive charge in CGT and excess negative charge in WS$_2$, leading to an interfacial electric dipole. The interfacial electric field and/or reduced electron density in the CGT could induce an ultrafast modification of the magnetic anisotropy. 
This corresponds to the transient anisotropy field $\Delta H_{int}$ in the LLG simulation, and its sign must be positive (i.e., favoring perpendicular magnetization) to account for the experimentally observed phase of $\phi_0$ near $+90 \degree$.
This electric-dipole-induced change in magnetic anisotropy can only happen in the WS$_2$/CGT heterostructure and leads to an opposite torque compared to CGT, where LLG modeling finds that the torque in CGT corresponds to a negative $\Delta H_{int}$.

Interestingly, the type-II band alignment would promote charge transfer in equilibrium as well, which influences $H_{int,0}$ with the same sign as $\Delta H_{int}$.
This would lead to an enhanced PMA in the heterostructure in equilibrium, which matches our observation in Fig.~\ref{fig:hysteresis_loop}. 
When the pump pulse is used to trigger magnetization dynamics, it promotes charge transfer even more to further enhance $H_{int}$, leading to the light-induced electric-dipole mechanism and precessional dynamics with larger amplitude.
Therefore, charge transfer in a type-II heterostructure provides a unified explanation for both the changes in PMA (equilibrium property) and in the pump-induced torque (dynamic property) when a WS$_2$ overlayer is added to CGT.

We investigate the band alignment with density functional theory (DFT) calculations and photocurrent measurements (see SM). For DFT, the sample consists of a monolayer of WS$_2$ on bulk-like CGT, which is appropriate for comparison with our samples of thickness $\sim$10 monolayers. For the band alignment, we find that a monolayer of WS$_2$ on CGT produces a type-II band alignment with the CGT bands (both valence band maxima and conduction band minima) at higher energy as compared to those of WS$_2$. In addition, DFT calculations show lower band gap (0.32 eV) of WS$_2$/CGT than that of CGT (0.6 eV) (Fig. \ref{fig:wavaelength}e, \ref{fig:wavaelength}f), confirming both magnetic semiconductors with Cr magnetic moments of 3.7 $\mu_B$ and 3.5 $\mu_B$, respectively.

We also investigate magnetic anisotropy by DFT and find enhanced PMA (E$_{110}$ - E$_{001}$) when monolayer WS$_2$ is placed on top of the CGT as compared to CGT. One electron excess positive charge in CGT enhances PMA by 5.2 meV in the WS$_2$/CGT heterostructure. In the case that the change in magnetic anisotropy is due to equilibrium charge transfer, then the enhanced PMA corresponds to an increased $H_{int,0}$ and a positive $\Delta H_{int}$ resulting in the observed dynamics.

\begin{figure}
\setlabel{pos=nw,fontsize=\scriptsize} 
  \subfloat{\xincludegraphics[width=0.32\textwidth]{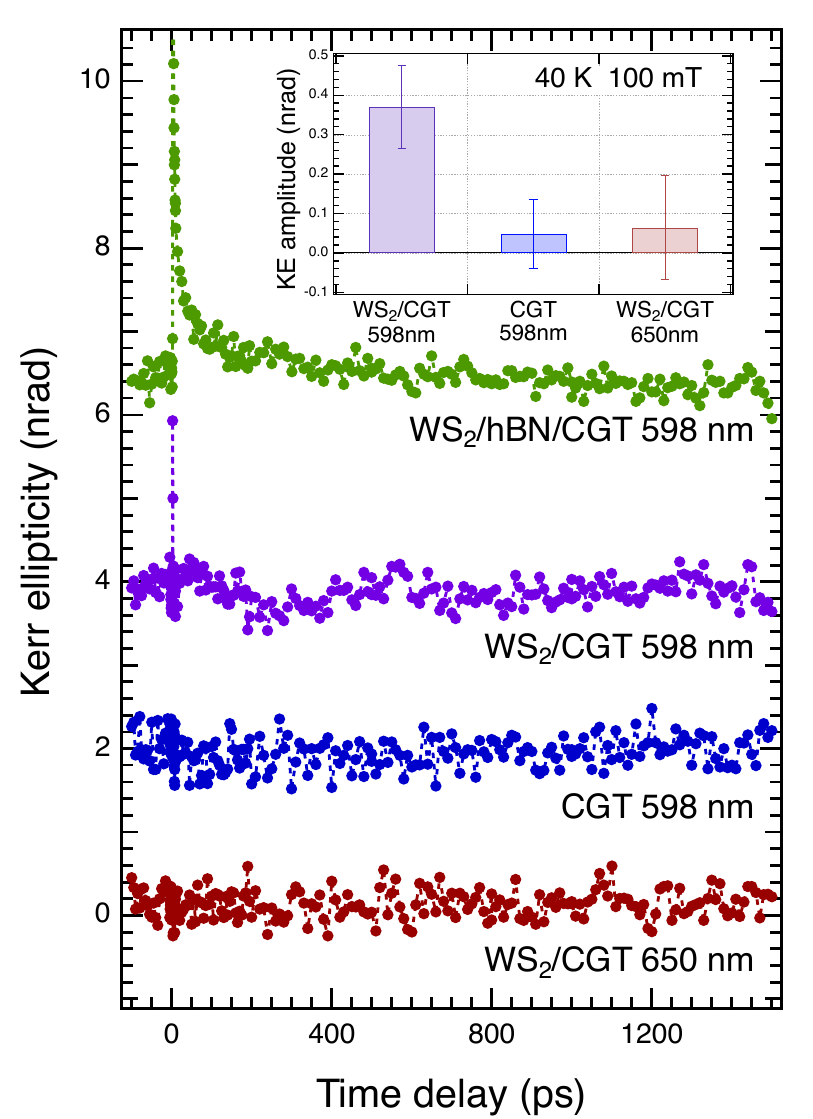}\label{fig:helicity_compare}}
  \hfill
  
\caption{\textbf{Pump helicity-dependent dynamics.} Helicity-moddulated pump induced spin dynamics at 40\,K with external field of 100\,mT in WS$_2$/hBN/CGT using 598\,nm pump (green), WS$_2$/CGT using 598\,nm pump (purple), CGT using 598\,nm pump (blue) and WS$_2$/CGT using 650\,nm (maroon). Inset: Extracted precession amplitude $a$ from the bottom three curves, indicating that the helicity-dependent spin selected dynamics becomes observable only when the WS$_2$ is resonantly pumped in WS$_2$/CGT region.
}\label{fig:helicity}
\end{figure}

Comparing with recent electrostatic gating experiments on CGT further supports this picture \cite{verzhbitskiy_controlling_2020,hendriks_electric_2024}. 
These experiments establish that applying gate voltages that reduce the electron density in CGT enhances the perpendicular magnetic anisotropy (PMA) \cite{weisheit_electric_2007, duan_surface_2008, nakamura_giant_2009, jiang_electrical_2018}. 
Our system operates via a similar principle: the lower energy bands of WS$_2$ induce a charge transfer to reduce the electron density in the underlying CGT (analogous to using $\alpha$-RuCl$_3$ for p-type doping of WSe$_2$ \cite{xie_low_2024, sternbach_quenched_2023}).
The experimentally observed PMA enhancement from placing WS$_2$ on CGT and the positive sign of $\Delta H_{int}$ for WS$_2$/CGT are both consistent with the gating results.
Although the specific interfacial details differ (e.g., the electric field distributions), the observation of similar trends nevertheless provides additional support.

To summarize, the proposed mechanism of a light-induced magnetoelectric effect in WS$_2$/CGT provides an elegant explanation that accounts for the experimental observations of large relative phase shift ($\sim$180$\degree$), enhanced precession amplitude compared to CGT (Fig.~\ref{fig:compare_precession_40K}) and the relative insensitivity to the pump wavelength (Fig.~\ref{fig:table}). The type-II band alignment in the heterostructure is supported by photocurrent measurements, and the proposed mechanism for anisotropy change based on charge transfer is consistent with established electrostatic gating results in CGT. DFT calculations support both the type-II alignment and the enhancement of of PMA when WS$_2$ is layered onto CGT. Numerical simulations show that the experimental TR-MOKE data and their field dependence can be well described by LLG modeling.

Lastly, we investigate the role of pump pulse helicity and the associated valley–spin polarization in WS$_2$ on driving spin dynamics in the adjacent  CGT layer. Through helicity-modulated pump pulse (measuring the difference between RCP and LCP pump), we eliminate thermally induced dynamics signal and exclusively resolve the precessional dynamics arising from spin-selective excitation by circularly polarized light. The results obtained in different regions of the sample are shown in Fig.~\ref{fig:helicity}. When WS$_2$ and CGT are separated by a thick hBN to block charge transfer (green curve), the sharp peak appearing immediately after the pump pulse arrives is attributed to polarized spin excited by circularly polarized pump in WS$_2$ that are unable to transfer into CGT. Interestingly, when the CGT is interfaced with WS$_2$, the peak is largely suppressed, while precessional dynamics from the transferred polarized spins emerge afterwards (purple curve). Using the same fitting procedure described previously, we obtain a phase $\phi_0$ of $\sim$62$\degree$ here as opposed to $\sim$74$\degree$ for the overall dynamics (Fig.\ref{fig:phi_vs_H_40K}). 
The shift of $\phi_0$ toward zero is consistent with the expectation that the transferred polarized spins exert an in-plane torque on the magnetization in CGT.
To rule out potential thermally-induced artifacts associated with the different absorption of RCP and LCP light in the CGT, we perform control experiments on CGT alone (blue data) and on WS$_2$/CGT at longer wavelength (maroon data), both of which promote absorption in the CGT. In both cases, no oscillations are observed, ruling out this potential artifact.

\section*{Acknowledgements}
\label{sec:acknowledgements}

The research was primarily supported as part of the Center for Energy Efficient Magnonics an Energy Frontier Research Center funded by the U.S. Department of Energy (DOE), Office of Science, Basic Energy Sciences (BES), under Award \# DE-AC02-76SF00515 (time-resolved magneto-optic Kerr effect, optical measurements, LLG simulations, DFT calculations) and under DOE BES Award No.~DE-SC0016379 (photocurrent measurements). 
S.S. acknowledges the financial support from U.S.~Office of Naval Research (ONR) under Award No.~N00014-23-1-2751, National Science Foundation (NSF) through Grant No.~DMR-2210510, and NSF-CAREER Award through Grant No.~ECCS-2339723 (sample fabrication).
J.K. acknowledges financial support from ONR under Award No.~N00014-23-1-2751 and NSF-CAREER Award under Grant No.~DMR-2339309 (sample fabrication).

The authors thank V\'ictor Hugo Ortiz Hern\'andez for assistance with Kittel equation analysis, and Zhenhong Cui and Raghvendra Posti for their assistance with heterostructure and device fabrication.

\bibliography{main.bib}

\begin{thebibliography}{55}%
\makeatletter
\providecommand \@ifxundefined [1]{%
 \@ifx{#1\undefined}
}%
\providecommand \@ifnum [1]{%
 \ifnum #1\expandafter \@firstoftwo
 \else \expandafter \@secondoftwo
 \fi
}%
\providecommand \@ifx [1]{%
 \ifx #1\expandafter \@firstoftwo
 \else \expandafter \@secondoftwo
 \fi
}%
\providecommand \natexlab [1]{#1}%
\providecommand \enquote  [1]{``#1''}%
\providecommand \bibnamefont  [1]{#1}%
\providecommand \bibfnamefont [1]{#1}%
\providecommand \citenamefont [1]{#1}%
\providecommand \href@noop [0]{\@secondoftwo}%
\providecommand \href [0]{\begingroup \@sanitize@url \@href}%
\providecommand \@href[1]{\@@startlink{#1}\@@href}%
\providecommand \@@href[1]{\endgroup#1\@@endlink}%
\providecommand \@sanitize@url [0]{\catcode `\\12\catcode `\$12\catcode `\&12\catcode `\#12\catcode `\^12\catcode `\_12\catcode `\%12\relax}%
\providecommand \@@startlink[1]{}%
\providecommand \@@endlink[0]{}%
\providecommand \url  [0]{\begingroup\@sanitize@url \@url }%
\providecommand \@url [1]{\endgroup\@href {#1}{\urlprefix }}%
\providecommand \urlprefix  [0]{URL }%
\providecommand \Eprint [0]{\href }%
\providecommand \doibase [0]{https://doi.org/}%
\providecommand \selectlanguage [0]{\@gobble}%
\providecommand \bibinfo  [0]{\@secondoftwo}%
\providecommand \bibfield  [0]{\@secondoftwo}%
\providecommand \translation [1]{[#1]}%
\providecommand \BibitemOpen [0]{}%
\providecommand \bibitemStop [0]{}%
\providecommand \bibitemNoStop [0]{.\EOS\space}%
\providecommand \EOS [0]{\spacefactor3000\relax}%
\providecommand \BibitemShut  [1]{\csname bibitem#1\endcsname}%
\let\auto@bib@innerbib\@empty
\bibitem [{\citenamefont {Kirilyuk}\ \emph {et~al.}(2010)\citenamefont {Kirilyuk}, \citenamefont {Kimel},\ and\ \citenamefont {Rasing}}]{kirilyuk_ultrafast_2010}%
  \BibitemOpen
  \bibfield  {author} {\bibinfo {author} {\bibfnamefont {A.}~\bibnamefont {Kirilyuk}}, \bibinfo {author} {\bibfnamefont {A.~V.}\ \bibnamefont {Kimel}},\ and\ \bibinfo {author} {\bibfnamefont {T.}~\bibnamefont {Rasing}},\ }\bibfield  {title} {\bibinfo {title} {Ultrafast optical manipulation of magnetic order},\ }\href {https://doi.org/10.1103/RevModPhys.82.2731} {\bibfield  {journal} {\bibinfo  {journal} {Reviews of Modern Physics}\ }\textbf {\bibinfo {volume} {82}},\ \bibinfo {pages} {2731} (\bibinfo {year} {2010})}\BibitemShut {NoStop}%
\bibitem [{\citenamefont {Dabrowski}\ \emph {et~al.}(2022)\citenamefont {Dabrowski}, \citenamefont {Guo}, \citenamefont {Strungaru}, \citenamefont {Keatley}, \citenamefont {Withers}, \citenamefont {Santos},\ and\ \citenamefont {Hicken}}]{dabrowski_all-optical_2022}%
  \BibitemOpen
  \bibfield  {author} {\bibinfo {author} {\bibfnamefont {M.}~\bibnamefont {Dabrowski}}, \bibinfo {author} {\bibfnamefont {S.}~\bibnamefont {Guo}}, \bibinfo {author} {\bibfnamefont {M.}~\bibnamefont {Strungaru}}, \bibinfo {author} {\bibfnamefont {P.~S.}\ \bibnamefont {Keatley}}, \bibinfo {author} {\bibfnamefont {F.}~\bibnamefont {Withers}}, \bibinfo {author} {\bibfnamefont {E.~J.~G.}\ \bibnamefont {Santos}},\ and\ \bibinfo {author} {\bibfnamefont {R.~J.}\ \bibnamefont {Hicken}},\ }\bibfield  {title} {\bibinfo {title} {All-optical control of spin in a {2D} van der {Waals} magnet},\ }\href {https://doi.org/10.1038/s41467-022-33343-4} {\bibfield  {journal} {\bibinfo  {journal} {Nature Communications}\ }\textbf {\bibinfo {volume} {13}},\ \bibinfo {pages} {5976} (\bibinfo {year} {2022})}\BibitemShut {NoStop}%
\bibitem [{\citenamefont {Zhang}\ \emph {et~al.}(2020{\natexlab{a}})\citenamefont {Zhang}, \citenamefont {Li}, \citenamefont {Weber}, \citenamefont {Goldberger}, \citenamefont {Mak},\ and\ \citenamefont {Shan}}]{zhang_gate-tunable_2020}%
  \BibitemOpen
  \bibfield  {author} {\bibinfo {author} {\bibfnamefont {X.-X.}\ \bibnamefont {Zhang}}, \bibinfo {author} {\bibfnamefont {L.}~\bibnamefont {Li}}, \bibinfo {author} {\bibfnamefont {D.}~\bibnamefont {Weber}}, \bibinfo {author} {\bibfnamefont {J.}~\bibnamefont {Goldberger}}, \bibinfo {author} {\bibfnamefont {K.~F.}\ \bibnamefont {Mak}},\ and\ \bibinfo {author} {\bibfnamefont {J.}~\bibnamefont {Shan}},\ }\bibfield  {title} {\bibinfo {title} {Gate-tunable spin waves in antiferromagnetic atomic bilayers},\ }\href {https://doi.org/10.1038/s41563-020-0713-9} {\bibfield  {journal} {\bibinfo  {journal} {Nature Materials}\ }\textbf {\bibinfo {volume} {19}},\ \bibinfo {pages} {838} (\bibinfo {year} {2020}{\natexlab{a}})}\BibitemShut {NoStop}%
\bibitem [{\citenamefont {Bae}\ \emph {et~al.}(2022)\citenamefont {Bae}, \citenamefont {Wang}, \citenamefont {Scheie}, \citenamefont {Xu}, \citenamefont {Chica}, \citenamefont {Diederich}, \citenamefont {Cenker}, \citenamefont {Ziebel}, \citenamefont {Bai}, \citenamefont {Ren}, \citenamefont {Dean}, \citenamefont {Delor}, \citenamefont {Xu}, \citenamefont {Roy}, \citenamefont {Kent},\ and\ \citenamefont {Zhu}}]{bae_exciton-coupled_2022}%
  \BibitemOpen
  \bibfield  {author} {\bibinfo {author} {\bibfnamefont {Y.~J.}\ \bibnamefont {Bae}}, \bibinfo {author} {\bibfnamefont {J.}~\bibnamefont {Wang}}, \bibinfo {author} {\bibfnamefont {A.}~\bibnamefont {Scheie}}, \bibinfo {author} {\bibfnamefont {J.}~\bibnamefont {Xu}}, \bibinfo {author} {\bibfnamefont {D.~G.}\ \bibnamefont {Chica}}, \bibinfo {author} {\bibfnamefont {G.~M.}\ \bibnamefont {Diederich}}, \bibinfo {author} {\bibfnamefont {J.}~\bibnamefont {Cenker}}, \bibinfo {author} {\bibfnamefont {M.~E.}\ \bibnamefont {Ziebel}}, \bibinfo {author} {\bibfnamefont {Y.}~\bibnamefont {Bai}}, \bibinfo {author} {\bibfnamefont {H.}~\bibnamefont {Ren}}, \bibinfo {author} {\bibfnamefont {C.~R.}\ \bibnamefont {Dean}}, \bibinfo {author} {\bibfnamefont {M.}~\bibnamefont {Delor}}, \bibinfo {author} {\bibfnamefont {X.}~\bibnamefont {Xu}}, \bibinfo {author} {\bibfnamefont {X.}~\bibnamefont {Roy}}, \bibinfo {author} {\bibfnamefont {A.~D.}\ \bibnamefont {Kent}},\ and\ \bibinfo {author} {\bibfnamefont {X.}~\bibnamefont {Zhu}},\
  }\bibfield  {title} {\bibinfo {title} {Exciton-coupled coherent magnons in a {2D} semiconductor},\ }\href {https://doi.org/10.1038/s41586-022-05024-1} {\bibfield  {journal} {\bibinfo  {journal} {Nature}\ }\textbf {\bibinfo {volume} {609}},\ \bibinfo {pages} {282} (\bibinfo {year} {2022})}\BibitemShut {NoStop}%
\bibitem [{\citenamefont {Diederich}\ \emph {et~al.}(2023)\citenamefont {Diederich}, \citenamefont {Cenker}, \citenamefont {Ren}, \citenamefont {Fonseca}, \citenamefont {Chica}, \citenamefont {Bae}, \citenamefont {Zhu}, \citenamefont {Roy}, \citenamefont {Cao}, \citenamefont {Xiao},\ and\ \citenamefont {Xu}}]{diederich_tunable_2023}%
  \BibitemOpen
  \bibfield  {author} {\bibinfo {author} {\bibfnamefont {G.~M.}\ \bibnamefont {Diederich}}, \bibinfo {author} {\bibfnamefont {J.}~\bibnamefont {Cenker}}, \bibinfo {author} {\bibfnamefont {Y.}~\bibnamefont {Ren}}, \bibinfo {author} {\bibfnamefont {J.}~\bibnamefont {Fonseca}}, \bibinfo {author} {\bibfnamefont {D.~G.}\ \bibnamefont {Chica}}, \bibinfo {author} {\bibfnamefont {Y.~J.}\ \bibnamefont {Bae}}, \bibinfo {author} {\bibfnamefont {X.}~\bibnamefont {Zhu}}, \bibinfo {author} {\bibfnamefont {X.}~\bibnamefont {Roy}}, \bibinfo {author} {\bibfnamefont {T.}~\bibnamefont {Cao}}, \bibinfo {author} {\bibfnamefont {D.}~\bibnamefont {Xiao}},\ and\ \bibinfo {author} {\bibfnamefont {X.}~\bibnamefont {Xu}},\ }\bibfield  {title} {\bibinfo {title} {Tunable interaction between excitons and hybridized magnons in a layered semiconductor},\ }\href {https://doi.org/10.1038/s41565-022-01259-1} {\bibfield  {journal} {\bibinfo  {journal} {Nature Nanotechnology}\ }\textbf {\bibinfo {volume} {18}},\ \bibinfo {pages} {23} (\bibinfo
  {year} {2023})}\BibitemShut {NoStop}%
\bibitem [{\citenamefont {Sun}\ \emph {et~al.}(2024)\citenamefont {Sun}, \citenamefont {Meng}, \citenamefont {Lee}, \citenamefont {Soll}, \citenamefont {Zhang}, \citenamefont {Ramesh}, \citenamefont {Yao}, \citenamefont {Sofer},\ and\ \citenamefont {Orenstein}}]{sun_dipolar_2024}%
  \BibitemOpen
  \bibfield  {author} {\bibinfo {author} {\bibfnamefont {Y.}~\bibnamefont {Sun}}, \bibinfo {author} {\bibfnamefont {F.}~\bibnamefont {Meng}}, \bibinfo {author} {\bibfnamefont {C.}~\bibnamefont {Lee}}, \bibinfo {author} {\bibfnamefont {A.}~\bibnamefont {Soll}}, \bibinfo {author} {\bibfnamefont {H.}~\bibnamefont {Zhang}}, \bibinfo {author} {\bibfnamefont {R.}~\bibnamefont {Ramesh}}, \bibinfo {author} {\bibfnamefont {J.}~\bibnamefont {Yao}}, \bibinfo {author} {\bibfnamefont {Z.}~\bibnamefont {Sofer}},\ and\ \bibinfo {author} {\bibfnamefont {J.}~\bibnamefont {Orenstein}},\ }\bibfield  {title} {\bibinfo {title} {Dipolar spin wave packet transport in a van der {Waals} antiferromagnet},\ }\href {https://doi.org/10.1038/s41567-024-02387-2} {\bibfield  {journal} {\bibinfo  {journal} {Nature Physics}\ }\textbf {\bibinfo {volume} {20}},\ \bibinfo {pages} {794} (\bibinfo {year} {2024})}\BibitemShut {NoStop}%
\bibitem [{\citenamefont {Bartram}\ \emph {et~al.}(2023)\citenamefont {Bartram}, \citenamefont {Li}, \citenamefont {Liu}, \citenamefont {Xu}, \citenamefont {Wang}, \citenamefont {Che}, \citenamefont {Li}, \citenamefont {Wu}, \citenamefont {Xu}, \citenamefont {Zhang}, \citenamefont {Yang},\ and\ \citenamefont {Yang}}]{bartram_real-time_2023}%
  \BibitemOpen
  \bibfield  {author} {\bibinfo {author} {\bibfnamefont {F.~M.}\ \bibnamefont {Bartram}}, \bibinfo {author} {\bibfnamefont {M.}~\bibnamefont {Li}}, \bibinfo {author} {\bibfnamefont {L.}~\bibnamefont {Liu}}, \bibinfo {author} {\bibfnamefont {Z.}~\bibnamefont {Xu}}, \bibinfo {author} {\bibfnamefont {Y.}~\bibnamefont {Wang}}, \bibinfo {author} {\bibfnamefont {M.}~\bibnamefont {Che}}, \bibinfo {author} {\bibfnamefont {H.}~\bibnamefont {Li}}, \bibinfo {author} {\bibfnamefont {Y.}~\bibnamefont {Wu}}, \bibinfo {author} {\bibfnamefont {Y.}~\bibnamefont {Xu}}, \bibinfo {author} {\bibfnamefont {J.}~\bibnamefont {Zhang}}, \bibinfo {author} {\bibfnamefont {S.}~\bibnamefont {Yang}},\ and\ \bibinfo {author} {\bibfnamefont {L.}~\bibnamefont {Yang}},\ }\bibfield  {title} {\bibinfo {title} {Real-time observation of magnetization and magnon dynamics in a two-dimensional topological antiferromagnet {MnBi}$_{\textrm{2}}${Te}$_{\textrm{4}}$},\ }\href {https://doi.org/10.1016/j.scib.2023.10.003} {\bibfield  {journal} {\bibinfo
  {journal} {Science Bulletin}\ }\textbf {\bibinfo {volume} {68}},\ \bibinfo {pages} {2734} (\bibinfo {year} {2023})}\BibitemShut {NoStop}%
\bibitem [{\citenamefont {Rongione}\ \emph {et~al.}(2023)\citenamefont {Rongione}, \citenamefont {Gueckstock}, \citenamefont {Mattern}, \citenamefont {Gomonay}, \citenamefont {Meer}, \citenamefont {Schmitt}, \citenamefont {Ramos}, \citenamefont {Kikkawa}, \citenamefont {Mi{\v{c}}ica}, \citenamefont {Saitoh}, \citenamefont {Sinova}, \citenamefont {Jaffr{\`e}s}, \citenamefont {Mangeney}, \citenamefont {Goennenwein}, \citenamefont {Gepr{\"a}gs}, \citenamefont {Kampfrath}, \citenamefont {Kl{\"a}ui}, \citenamefont {Bargheer}, \citenamefont {Seifert}, \citenamefont {Dhillon},\ and\ \citenamefont {Lebrun}}]{rongione_emission_2023}%
  \BibitemOpen
  \bibfield  {author} {\bibinfo {author} {\bibfnamefont {E.}~\bibnamefont {Rongione}}, \bibinfo {author} {\bibfnamefont {O.}~\bibnamefont {Gueckstock}}, \bibinfo {author} {\bibfnamefont {M.}~\bibnamefont {Mattern}}, \bibinfo {author} {\bibfnamefont {O.}~\bibnamefont {Gomonay}}, \bibinfo {author} {\bibfnamefont {H.}~\bibnamefont {Meer}}, \bibinfo {author} {\bibfnamefont {C.}~\bibnamefont {Schmitt}}, \bibinfo {author} {\bibfnamefont {R.}~\bibnamefont {Ramos}}, \bibinfo {author} {\bibfnamefont {T.}~\bibnamefont {Kikkawa}}, \bibinfo {author} {\bibfnamefont {M.}~\bibnamefont {Mi{\v{c}}ica}}, \bibinfo {author} {\bibfnamefont {E.}~\bibnamefont {Saitoh}}, \bibinfo {author} {\bibfnamefont {J.}~\bibnamefont {Sinova}}, \bibinfo {author} {\bibfnamefont {H.}~\bibnamefont {Jaffr{\`e}s}}, \bibinfo {author} {\bibfnamefont {J.}~\bibnamefont {Mangeney}}, \bibinfo {author} {\bibfnamefont {S.~T.~B.}\ \bibnamefont {Goennenwein}}, \bibinfo {author} {\bibfnamefont {S.}~\bibnamefont {Gepr{\"a}gs}}, \bibinfo {author} {\bibfnamefont
  {T.}~\bibnamefont {Kampfrath}}, \bibinfo {author} {\bibfnamefont {M.}~\bibnamefont {Kl{\"a}ui}}, \bibinfo {author} {\bibfnamefont {M.}~\bibnamefont {Bargheer}}, \bibinfo {author} {\bibfnamefont {T.~S.}\ \bibnamefont {Seifert}}, \bibinfo {author} {\bibfnamefont {S.}~\bibnamefont {Dhillon}},\ and\ \bibinfo {author} {\bibfnamefont {R.}~\bibnamefont {Lebrun}},\ }\bibfield  {title} {\bibinfo {title} {Emission of coherent {THz} magnons in an antiferromagnetic insulator triggered by ultrafast spin--phonon interactions},\ }\href {https://doi.org/10.1038/s41467-023-37509-6} {\bibfield  {journal} {\bibinfo  {journal} {Nature Communications}\ }\textbf {\bibinfo {volume} {14}},\ \bibinfo {pages} {1818} (\bibinfo {year} {2023})}\BibitemShut {NoStop}%
\bibitem [{\citenamefont {Li}\ \emph {et~al.}(2025)\citenamefont {Li}, \citenamefont {Wu}, \citenamefont {Liu}, \citenamefont {Cheng}, \citenamefont {Gong}, \citenamefont {Tong}, \citenamefont {Liu}, \citenamefont {He}, \citenamefont {Xiu}, \citenamefont {Zhao}, \citenamefont {Meng},\ and\ \citenamefont {Wu}}]{li_above-curie-temperature_2025}%
  \BibitemOpen
  \bibfield  {author} {\bibinfo {author} {\bibfnamefont {P.}~\bibnamefont {Li}}, \bibinfo {author} {\bibfnamefont {N.}~\bibnamefont {Wu}}, \bibinfo {author} {\bibfnamefont {S.}~\bibnamefont {Liu}}, \bibinfo {author} {\bibfnamefont {Y.}~\bibnamefont {Cheng}}, \bibinfo {author} {\bibfnamefont {P.}~\bibnamefont {Gong}}, \bibinfo {author} {\bibfnamefont {J.}~\bibnamefont {Tong}}, \bibinfo {author} {\bibfnamefont {J.}~\bibnamefont {Liu}}, \bibinfo {author} {\bibfnamefont {W.}~\bibnamefont {He}}, \bibinfo {author} {\bibfnamefont {F.}~\bibnamefont {Xiu}}, \bibinfo {author} {\bibfnamefont {J.}~\bibnamefont {Zhao}}, \bibinfo {author} {\bibfnamefont {S.}~\bibnamefont {Meng}},\ and\ \bibinfo {author} {\bibfnamefont {X.}~\bibnamefont {Wu}},\ }\bibfield  {title} {\bibinfo {title} {Above-{Curie}-temperature ultrafast terahertz emission and spin current generation in a {2D} superlattice ({Fe}$_{\textrm{3}}${GeTe}$_{\textrm{2}}$/{CrSb})$_{\textrm{3}}$},\ }\href {https://doi.org/10.1093/nsr/nwae447} {\bibfield  {journal}
  {\bibinfo  {journal} {National Science Review}\ }\textbf {\bibinfo {volume} {12}},\ \bibinfo {pages} {nwae447} (\bibinfo {year} {2025})}\BibitemShut {NoStop}%
\bibitem [{\citenamefont {Kimel}\ and\ \citenamefont {Li}(2019)}]{kimel_writing_2019}%
  \BibitemOpen
  \bibfield  {author} {\bibinfo {author} {\bibfnamefont {A.~V.}\ \bibnamefont {Kimel}}\ and\ \bibinfo {author} {\bibfnamefont {M.}~\bibnamefont {Li}},\ }\bibfield  {title} {\bibinfo {title} {Writing magnetic memory with ultrashort light pulses},\ }\href {https://doi.org/10.1038/s41578-019-0086-3} {\bibfield  {journal} {\bibinfo  {journal} {Nature Reviews Materials}\ }\textbf {\bibinfo {volume} {4}},\ \bibinfo {pages} {189} (\bibinfo {year} {2019})}\BibitemShut {NoStop}%
\bibitem [{\citenamefont {Barman}\ \emph {et~al.}(2021)\citenamefont {Barman}, \citenamefont {Gubbiotti}, \citenamefont {Ladak}, \citenamefont {Adeyeye}, \citenamefont {Krawczyk}, \citenamefont {Gr{\"a}fe}, \citenamefont {Adelmann}, \citenamefont {Cotofana}, \citenamefont {Naeemi}, \citenamefont {Vasyuchka}, \citenamefont {Hillebrands}, \citenamefont {Nikitov}, \citenamefont {Yu}, \citenamefont {Grundler}, \citenamefont {Sadovnikov}, \citenamefont {Grachev}, \citenamefont {Sheshukova}, \citenamefont {Duquesne}, \citenamefont {Marangolo}, \citenamefont {Csaba}, \citenamefont {Porod}, \citenamefont {Demidov}, \citenamefont {Urazhdin}, \citenamefont {Demokritov}, \citenamefont {Albisetti}, \citenamefont {Petti}, \citenamefont {Bertacco}, \citenamefont {Schultheiss}, \citenamefont {Kruglyak}, \citenamefont {Poimanov}, \citenamefont {Sahoo}, \citenamefont {Sinha}, \citenamefont {Yang}, \citenamefont {M{\"u}nzenberg}, \citenamefont {Moriyama}, \citenamefont {Mizukami}, \citenamefont {Landeros}, \citenamefont
  {Gallardo}, \citenamefont {Carlotti}, \citenamefont {Kim}, \citenamefont {Stamps}, \citenamefont {Camley}, \citenamefont {Rana}, \citenamefont {Otani}, \citenamefont {Yu}, \citenamefont {Yu}, \citenamefont {Bauer}, \citenamefont {Back}, \citenamefont {Uhrig}, \citenamefont {Dobrovolskiy}, \citenamefont {Budinska}, \citenamefont {Qin}, \citenamefont {van Dijken}, \citenamefont {Chumak}, \citenamefont {Khitun}, \citenamefont {Nikonov}, \citenamefont {Young}, \citenamefont {Zingsem},\ and\ \citenamefont {Winklhofer}}]{barman_2021_2021}%
  \BibitemOpen
  \bibfield  {author} {\bibinfo {author} {\bibfnamefont {A.}~\bibnamefont {Barman}}, \bibinfo {author} {\bibfnamefont {G.}~\bibnamefont {Gubbiotti}}, \bibinfo {author} {\bibfnamefont {S.}~\bibnamefont {Ladak}}, \bibinfo {author} {\bibfnamefont {A.~O.}\ \bibnamefont {Adeyeye}}, \bibinfo {author} {\bibfnamefont {M.}~\bibnamefont {Krawczyk}}, \bibinfo {author} {\bibfnamefont {J.}~\bibnamefont {Gr{\"a}fe}}, \bibinfo {author} {\bibfnamefont {C.}~\bibnamefont {Adelmann}}, \bibinfo {author} {\bibfnamefont {S.}~\bibnamefont {Cotofana}}, \bibinfo {author} {\bibfnamefont {A.}~\bibnamefont {Naeemi}}, \bibinfo {author} {\bibfnamefont {V.~I.}\ \bibnamefont {Vasyuchka}}, \bibinfo {author} {\bibfnamefont {B.}~\bibnamefont {Hillebrands}}, \bibinfo {author} {\bibfnamefont {S.~A.}\ \bibnamefont {Nikitov}}, \bibinfo {author} {\bibfnamefont {H.}~\bibnamefont {Yu}}, \bibinfo {author} {\bibfnamefont {D.}~\bibnamefont {Grundler}}, \bibinfo {author} {\bibfnamefont {A.~V.}\ \bibnamefont {Sadovnikov}}, \bibinfo {author} {\bibfnamefont
  {A.~A.}\ \bibnamefont {Grachev}}, \bibinfo {author} {\bibfnamefont {S.~E.}\ \bibnamefont {Sheshukova}}, \bibinfo {author} {\bibfnamefont {J.-Y.}\ \bibnamefont {Duquesne}}, \bibinfo {author} {\bibfnamefont {M.}~\bibnamefont {Marangolo}}, \bibinfo {author} {\bibfnamefont {G.}~\bibnamefont {Csaba}}, \bibinfo {author} {\bibfnamefont {W.}~\bibnamefont {Porod}}, \bibinfo {author} {\bibfnamefont {V.~E.}\ \bibnamefont {Demidov}}, \bibinfo {author} {\bibfnamefont {S.}~\bibnamefont {Urazhdin}}, \bibinfo {author} {\bibfnamefont {S.~O.}\ \bibnamefont {Demokritov}}, \bibinfo {author} {\bibfnamefont {E.}~\bibnamefont {Albisetti}}, \bibinfo {author} {\bibfnamefont {D.}~\bibnamefont {Petti}}, \bibinfo {author} {\bibfnamefont {R.}~\bibnamefont {Bertacco}}, \bibinfo {author} {\bibfnamefont {H.}~\bibnamefont {Schultheiss}}, \bibinfo {author} {\bibfnamefont {V.~V.}\ \bibnamefont {Kruglyak}}, \bibinfo {author} {\bibfnamefont {V.~D.}\ \bibnamefont {Poimanov}}, \bibinfo {author} {\bibfnamefont {S.}~\bibnamefont {Sahoo}}, \bibinfo
  {author} {\bibfnamefont {J.}~\bibnamefont {Sinha}}, \bibinfo {author} {\bibfnamefont {H.}~\bibnamefont {Yang}}, \bibinfo {author} {\bibfnamefont {M.}~\bibnamefont {M{\"u}nzenberg}}, \bibinfo {author} {\bibfnamefont {T.}~\bibnamefont {Moriyama}}, \bibinfo {author} {\bibfnamefont {S.}~\bibnamefont {Mizukami}}, \bibinfo {author} {\bibfnamefont {P.}~\bibnamefont {Landeros}}, \bibinfo {author} {\bibfnamefont {R.~A.}\ \bibnamefont {Gallardo}}, \bibinfo {author} {\bibfnamefont {G.}~\bibnamefont {Carlotti}}, \bibinfo {author} {\bibfnamefont {J.-V.}\ \bibnamefont {Kim}}, \bibinfo {author} {\bibfnamefont {R.~L.}\ \bibnamefont {Stamps}}, \bibinfo {author} {\bibfnamefont {R.~E.}\ \bibnamefont {Camley}}, \bibinfo {author} {\bibfnamefont {B.}~\bibnamefont {Rana}}, \bibinfo {author} {\bibfnamefont {Y.}~\bibnamefont {Otani}}, \bibinfo {author} {\bibfnamefont {W.}~\bibnamefont {Yu}}, \bibinfo {author} {\bibfnamefont {T.}~\bibnamefont {Yu}}, \bibinfo {author} {\bibfnamefont {G.~E.~W.}\ \bibnamefont {Bauer}}, \bibinfo
  {author} {\bibfnamefont {C.}~\bibnamefont {Back}}, \bibinfo {author} {\bibfnamefont {G.~S.}\ \bibnamefont {Uhrig}}, \bibinfo {author} {\bibfnamefont {O.~V.}\ \bibnamefont {Dobrovolskiy}}, \bibinfo {author} {\bibfnamefont {B.}~\bibnamefont {Budinska}}, \bibinfo {author} {\bibfnamefont {H.}~\bibnamefont {Qin}}, \bibinfo {author} {\bibfnamefont {S.}~\bibnamefont {van Dijken}}, \bibinfo {author} {\bibfnamefont {A.~V.}\ \bibnamefont {Chumak}}, \bibinfo {author} {\bibfnamefont {A.}~\bibnamefont {Khitun}}, \bibinfo {author} {\bibfnamefont {D.~E.}\ \bibnamefont {Nikonov}}, \bibinfo {author} {\bibfnamefont {I.~A.}\ \bibnamefont {Young}}, \bibinfo {author} {\bibfnamefont {B.~W.}\ \bibnamefont {Zingsem}},\ and\ \bibinfo {author} {\bibfnamefont {M.}~\bibnamefont {Winklhofer}},\ }\bibfield  {title} {\bibinfo {title} {The 2021 magnonics roadmap},\ }\href {https://doi.org/10.1088/1361-648X/abec1a} {\bibfield  {journal} {\bibinfo  {journal} {Journal of Physics: Condensed Matter}\ }\textbf {\bibinfo {volume} {33}},\
  \bibinfo {pages} {413001} (\bibinfo {year} {2021})}\BibitemShut {NoStop}%
\bibitem [{\citenamefont {Beaurepaire}\ \emph {et~al.}(1996)\citenamefont {Beaurepaire}, \citenamefont {Merle}, \citenamefont {Daunois},\ and\ \citenamefont {Bigot}}]{beaurepaire_ultrafast_1996}%
  \BibitemOpen
  \bibfield  {author} {\bibinfo {author} {\bibfnamefont {E.}~\bibnamefont {Beaurepaire}}, \bibinfo {author} {\bibfnamefont {J.-C.}\ \bibnamefont {Merle}}, \bibinfo {author} {\bibfnamefont {A.}~\bibnamefont {Daunois}},\ and\ \bibinfo {author} {\bibfnamefont {J.-Y.}\ \bibnamefont {Bigot}},\ }\bibfield  {title} {\bibinfo {title} {Ultrafast spin dynamics in ferromagnetic nickel},\ }\href {https://doi.org/10.1103/PhysRevLett.76.4250} {\bibfield  {journal} {\bibinfo  {journal} {Physical Review Letters}\ }\textbf {\bibinfo {volume} {76}},\ \bibinfo {pages} {4250} (\bibinfo {year} {1996})}\BibitemShut {NoStop}%
\bibitem [{\citenamefont {Iihama}\ \emph {et~al.}(2014)\citenamefont {Iihama}, \citenamefont {Mizukami}, \citenamefont {Naganuma}, \citenamefont {Oogane}, \citenamefont {Ando},\ and\ \citenamefont {Miyazaki}}]{iihama_gilbert_2014}%
  \BibitemOpen
  \bibfield  {author} {\bibinfo {author} {\bibfnamefont {S.}~\bibnamefont {Iihama}}, \bibinfo {author} {\bibfnamefont {S.}~\bibnamefont {Mizukami}}, \bibinfo {author} {\bibfnamefont {H.}~\bibnamefont {Naganuma}}, \bibinfo {author} {\bibfnamefont {M.}~\bibnamefont {Oogane}}, \bibinfo {author} {\bibfnamefont {Y.}~\bibnamefont {Ando}},\ and\ \bibinfo {author} {\bibfnamefont {T.}~\bibnamefont {Miyazaki}},\ }\bibfield  {title} {\bibinfo {title} {Gilbert damping constants of {Ta}/{CoFeB}/{MgO}({Ta}) thin films measured by optical detection of precessional magnetization dynamics},\ }\href {https://doi.org/10.1103/PhysRevB.89.174416} {\bibfield  {journal} {\bibinfo  {journal} {Physical Review B}\ }\textbf {\bibinfo {volume} {89}},\ \bibinfo {pages} {174416} (\bibinfo {year} {2014})}\BibitemShut {NoStop}%
\bibitem [{\citenamefont {Tang}\ \emph {et~al.}(2023)\citenamefont {Tang}, \citenamefont {Alahmed}, \citenamefont {Mahdi}, \citenamefont {Xiong}, \citenamefont {Inman}, \citenamefont {McLaughlin}, \citenamefont {Zollitsch}, \citenamefont {Kim}, \citenamefont {Du}, \citenamefont {Kurebayashi}, \citenamefont {Santos}, \citenamefont {Zhang}, \citenamefont {Li},\ and\ \citenamefont {Jin}}]{tang_spin_2023}%
  \BibitemOpen
  \bibfield  {author} {\bibinfo {author} {\bibfnamefont {C.}~\bibnamefont {Tang}}, \bibinfo {author} {\bibfnamefont {L.}~\bibnamefont {Alahmed}}, \bibinfo {author} {\bibfnamefont {M.}~\bibnamefont {Mahdi}}, \bibinfo {author} {\bibfnamefont {Y.}~\bibnamefont {Xiong}}, \bibinfo {author} {\bibfnamefont {J.}~\bibnamefont {Inman}}, \bibinfo {author} {\bibfnamefont {N.~J.}\ \bibnamefont {McLaughlin}}, \bibinfo {author} {\bibfnamefont {C.}~\bibnamefont {Zollitsch}}, \bibinfo {author} {\bibfnamefont {T.~H.}\ \bibnamefont {Kim}}, \bibinfo {author} {\bibfnamefont {C.~R.}\ \bibnamefont {Du}}, \bibinfo {author} {\bibfnamefont {H.}~\bibnamefont {Kurebayashi}}, \bibinfo {author} {\bibfnamefont {E.~J.}\ \bibnamefont {Santos}}, \bibinfo {author} {\bibfnamefont {W.}~\bibnamefont {Zhang}}, \bibinfo {author} {\bibfnamefont {P.}~\bibnamefont {Li}},\ and\ \bibinfo {author} {\bibfnamefont {W.}~\bibnamefont {Jin}},\ }\bibfield  {title} {\bibinfo {title} {Spin dynamics in van der {Waals} magnetic systems},\ }\href
  {https://doi.org/10.1016/j.physrep.2023.09.002} {\bibfield  {journal} {\bibinfo  {journal} {Physics Reports}\ }\textbf {\bibinfo {volume} {1032}},\ \bibinfo {pages} {1} (\bibinfo {year} {2023})}\BibitemShut {NoStop}%
\bibitem [{\citenamefont {Zhang}\ \emph {et~al.}(2020{\natexlab{b}})\citenamefont {Zhang}, \citenamefont {Chen}, \citenamefont {Li}, \citenamefont {Guo}, \citenamefont {Wang}, \citenamefont {Han}, \citenamefont {He},\ and\ \citenamefont {Zhang}}]{zhang_laser-induced_2020}%
  \BibitemOpen
  \bibfield  {author} {\bibinfo {author} {\bibfnamefont {T.}~\bibnamefont {Zhang}}, \bibinfo {author} {\bibfnamefont {Y.}~\bibnamefont {Chen}}, \bibinfo {author} {\bibfnamefont {Y.}~\bibnamefont {Li}}, \bibinfo {author} {\bibfnamefont {Z.}~\bibnamefont {Guo}}, \bibinfo {author} {\bibfnamefont {Z.}~\bibnamefont {Wang}}, \bibinfo {author} {\bibfnamefont {Z.}~\bibnamefont {Han}}, \bibinfo {author} {\bibfnamefont {W.}~\bibnamefont {He}},\ and\ \bibinfo {author} {\bibfnamefont {J.}~\bibnamefont {Zhang}},\ }\bibfield  {title} {\bibinfo {title} {Laser-induced magnetization dynamics in a van der {Waals} ferromagnetic {Cr$_2$Ge$_2$Te$_6$} nanoflake},\ }\href {https://doi.org/10.1063/5.0006080} {\bibfield  {journal} {\bibinfo  {journal} {Applied Physics Letters}\ }\textbf {\bibinfo {volume} {116}},\ \bibinfo {pages} {223103} (\bibinfo {year} {2020}{\natexlab{b}})}\BibitemShut {NoStop}%
\bibitem [{\citenamefont {Hendriks}\ \emph {et~al.}(2024)\citenamefont {Hendriks}, \citenamefont {Rojas-Lopez}, \citenamefont {Koopmans},\ and\ \citenamefont {Guimaraes}}]{hendriks_electric_2024}%
  \BibitemOpen
  \bibfield  {author} {\bibinfo {author} {\bibfnamefont {F.}~\bibnamefont {Hendriks}}, \bibinfo {author} {\bibfnamefont {R.~R.}\ \bibnamefont {Rojas-Lopez}}, \bibinfo {author} {\bibfnamefont {B.}~\bibnamefont {Koopmans}},\ and\ \bibinfo {author} {\bibfnamefont {M.~H.~D.}\ \bibnamefont {Guimaraes}},\ }\bibfield  {title} {\bibinfo {title} {Electric control of optically-induced magnetization dynamics in a van der {Waals} ferromagnetic semiconductor},\ }\href {https://doi.org/10.1038/s41467-024-45623-2} {\bibfield  {journal} {\bibinfo  {journal} {Nature Communications}\ }\textbf {\bibinfo {volume} {15}},\ \bibinfo {pages} {1298} (\bibinfo {year} {2024})}\BibitemShut {NoStop}%
\bibitem [{\citenamefont {Kimel}\ \emph {et~al.}(2005)\citenamefont {Kimel}, \citenamefont {Kirilyuk}, \citenamefont {Usachev}, \citenamefont {Pisarev}, \citenamefont {Balbashov},\ and\ \citenamefont {Rasing}}]{kimel_ultrafast_2005}%
  \BibitemOpen
  \bibfield  {author} {\bibinfo {author} {\bibfnamefont {A.~V.}\ \bibnamefont {Kimel}}, \bibinfo {author} {\bibfnamefont {A.}~\bibnamefont {Kirilyuk}}, \bibinfo {author} {\bibfnamefont {P.~A.}\ \bibnamefont {Usachev}}, \bibinfo {author} {\bibfnamefont {R.~V.}\ \bibnamefont {Pisarev}}, \bibinfo {author} {\bibfnamefont {A.~M.}\ \bibnamefont {Balbashov}},\ and\ \bibinfo {author} {\bibfnamefont {T.}~\bibnamefont {Rasing}},\ }\bibfield  {title} {\bibinfo {title} {Ultrafast non-thermal control of magnetization by instantaneous photomagnetic pulses},\ }\href {https://doi.org/10.1038/nature03564} {\bibfield  {journal} {\bibinfo  {journal} {Nature}\ }\textbf {\bibinfo {volume} {435}},\ \bibinfo {pages} {655} (\bibinfo {year} {2005})}\BibitemShut {NoStop}%
\bibitem [{\citenamefont {Choi}\ \emph {et~al.}(2017)\citenamefont {Choi}, \citenamefont {Schleife},\ and\ \citenamefont {Cahill}}]{choi_optical-helicity-driven_2017}%
  \BibitemOpen
  \bibfield  {author} {\bibinfo {author} {\bibfnamefont {G.-M.}\ \bibnamefont {Choi}}, \bibinfo {author} {\bibfnamefont {A.}~\bibnamefont {Schleife}},\ and\ \bibinfo {author} {\bibfnamefont {D.~G.}\ \bibnamefont {Cahill}},\ }\bibfield  {title} {\bibinfo {title} {Optical-helicity-driven magnetization dynamics in metallic ferromagnets},\ }\href {https://doi.org/10.1038/ncomms15085} {\bibfield  {journal} {\bibinfo  {journal} {Nature Communications}\ }\textbf {\bibinfo {volume} {8}},\ \bibinfo {pages} {15085} (\bibinfo {year} {2017})}\BibitemShut {NoStop}%
\bibitem [{\citenamefont {Chernikov}\ \emph {et~al.}(2014)\citenamefont {Chernikov}, \citenamefont {Berkelbach}, \citenamefont {Hill}, \citenamefont {Rigosi}, \citenamefont {Li}, \citenamefont {Aslan}, \citenamefont {Reichman}, \citenamefont {Hybertsen},\ and\ \citenamefont {Heinz}}]{chernikov_exciton_2014}%
  \BibitemOpen
  \bibfield  {author} {\bibinfo {author} {\bibfnamefont {A.}~\bibnamefont {Chernikov}}, \bibinfo {author} {\bibfnamefont {T.~C.}\ \bibnamefont {Berkelbach}}, \bibinfo {author} {\bibfnamefont {H.~M.}\ \bibnamefont {Hill}}, \bibinfo {author} {\bibfnamefont {A.}~\bibnamefont {Rigosi}}, \bibinfo {author} {\bibfnamefont {Y.}~\bibnamefont {Li}}, \bibinfo {author} {\bibfnamefont {O.~B.}\ \bibnamefont {Aslan}}, \bibinfo {author} {\bibfnamefont {D.~R.}\ \bibnamefont {Reichman}}, \bibinfo {author} {\bibfnamefont {M.~S.}\ \bibnamefont {Hybertsen}},\ and\ \bibinfo {author} {\bibfnamefont {T.~F.}\ \bibnamefont {Heinz}},\ }\bibfield  {title} {\bibinfo {title} {Exciton binding energy and nonhydrogenic {Rydberg} series in monolayer {WS}$_{\textrm{2}}$},\ }\href {https://doi.org/10.1103/PhysRevLett.113.076802} {\bibfield  {journal} {\bibinfo  {journal} {Physical Review Letters}\ }\textbf {\bibinfo {volume} {113}},\ \bibinfo {pages} {076802} (\bibinfo {year} {2014})}\BibitemShut {NoStop}%
\bibitem [{\citenamefont {Mak}\ and\ \citenamefont {Shan}(2016)}]{mak_photonics_2016}%
  \BibitemOpen
  \bibfield  {author} {\bibinfo {author} {\bibfnamefont {K.~F.}\ \bibnamefont {Mak}}\ and\ \bibinfo {author} {\bibfnamefont {J.}~\bibnamefont {Shan}},\ }\bibfield  {title} {\bibinfo {title} {Photonics and optoelectronics of 2{D} semiconductor transition metal dichalcogenides},\ }\href {https://doi.org/10.1038/nphoton.2015.282} {\bibfield  {journal} {\bibinfo  {journal} {Nature Photonics}\ }\textbf {\bibinfo {volume} {10}},\ \bibinfo {pages} {216} (\bibinfo {year} {2016})}\BibitemShut {NoStop}%
\bibitem [{\citenamefont {Aivazian}\ \emph {et~al.}(2015)\citenamefont {Aivazian}, \citenamefont {Gong}, \citenamefont {Jones}, \citenamefont {Chu}, \citenamefont {Yan}, \citenamefont {Mandrus}, \citenamefont {Zhang}, \citenamefont {Cobden}, \citenamefont {Yao},\ and\ \citenamefont {Xu}}]{aivazian_magnetic_2015}%
  \BibitemOpen
  \bibfield  {author} {\bibinfo {author} {\bibfnamefont {G.}~\bibnamefont {Aivazian}}, \bibinfo {author} {\bibfnamefont {Z.}~\bibnamefont {Gong}}, \bibinfo {author} {\bibfnamefont {A.~M.}\ \bibnamefont {Jones}}, \bibinfo {author} {\bibfnamefont {R.-L.}\ \bibnamefont {Chu}}, \bibinfo {author} {\bibfnamefont {J.}~\bibnamefont {Yan}}, \bibinfo {author} {\bibfnamefont {D.~G.}\ \bibnamefont {Mandrus}}, \bibinfo {author} {\bibfnamefont {C.}~\bibnamefont {Zhang}}, \bibinfo {author} {\bibfnamefont {D.}~\bibnamefont {Cobden}}, \bibinfo {author} {\bibfnamefont {W.}~\bibnamefont {Yao}},\ and\ \bibinfo {author} {\bibfnamefont {X.}~\bibnamefont {Xu}},\ }\bibfield  {title} {\bibinfo {title} {Magnetic control of valley pseudospin in monolayer {WSe$_2$}},\ }\href {https://doi.org/10.1038/nphys3201} {\bibfield  {journal} {\bibinfo  {journal} {Nature Physics}\ }\textbf {\bibinfo {volume} {11}},\ \bibinfo {pages} {148} (\bibinfo {year} {2015})}\BibitemShut {NoStop}%
\bibitem [{\citenamefont {Zhong}\ \emph {et~al.}(2017)\citenamefont {Zhong}, \citenamefont {Seyler}, \citenamefont {Linpeng}, \citenamefont {Cheng}, \citenamefont {Sivadas}, \citenamefont {Huang}, \citenamefont {Schmidgall}, \citenamefont {Taniguchi}, \citenamefont {Watanabe}, \citenamefont {McGuire}, \citenamefont {Yao}, \citenamefont {Xiao}, \citenamefont {Fu},\ and\ \citenamefont {Xu}}]{zhong_van_2017}%
  \BibitemOpen
  \bibfield  {author} {\bibinfo {author} {\bibfnamefont {D.}~\bibnamefont {Zhong}}, \bibinfo {author} {\bibfnamefont {K.~L.}\ \bibnamefont {Seyler}}, \bibinfo {author} {\bibfnamefont {X.}~\bibnamefont {Linpeng}}, \bibinfo {author} {\bibfnamefont {R.}~\bibnamefont {Cheng}}, \bibinfo {author} {\bibfnamefont {N.}~\bibnamefont {Sivadas}}, \bibinfo {author} {\bibfnamefont {B.}~\bibnamefont {Huang}}, \bibinfo {author} {\bibfnamefont {E.}~\bibnamefont {Schmidgall}}, \bibinfo {author} {\bibfnamefont {T.}~\bibnamefont {Taniguchi}}, \bibinfo {author} {\bibfnamefont {K.}~\bibnamefont {Watanabe}}, \bibinfo {author} {\bibfnamefont {M.~A.}\ \bibnamefont {McGuire}}, \bibinfo {author} {\bibfnamefont {W.}~\bibnamefont {Yao}}, \bibinfo {author} {\bibfnamefont {D.}~\bibnamefont {Xiao}}, \bibinfo {author} {\bibfnamefont {K.-M.~C.}\ \bibnamefont {Fu}},\ and\ \bibinfo {author} {\bibfnamefont {X.}~\bibnamefont {Xu}},\ }\bibfield  {title} {\bibinfo {title} {Van der {Waals} engineering of ferromagnetic semiconductor heterostructures
  for spin and valleytronics},\ }\href {https://doi.org/10.1126/sciadv.1603113} {\bibfield  {journal} {\bibinfo  {journal} {Science Advances}\ }\textbf {\bibinfo {volume} {3}},\ \bibinfo {pages} {e1603113} (\bibinfo {year} {2017})}\BibitemShut {NoStop}%
\bibitem [{\citenamefont {Norden}\ \emph {et~al.}(2019)\citenamefont {Norden}, \citenamefont {Zhao}, \citenamefont {Zhang}, \citenamefont {Sabirianov}, \citenamefont {Petrou},\ and\ \citenamefont {Zeng}}]{norden_giant_2019}%
  \BibitemOpen
  \bibfield  {author} {\bibinfo {author} {\bibfnamefont {T.}~\bibnamefont {Norden}}, \bibinfo {author} {\bibfnamefont {C.}~\bibnamefont {Zhao}}, \bibinfo {author} {\bibfnamefont {P.}~\bibnamefont {Zhang}}, \bibinfo {author} {\bibfnamefont {R.}~\bibnamefont {Sabirianov}}, \bibinfo {author} {\bibfnamefont {A.}~\bibnamefont {Petrou}},\ and\ \bibinfo {author} {\bibfnamefont {H.}~\bibnamefont {Zeng}},\ }\bibfield  {title} {\bibinfo {title} {Giant valley splitting in monolayer {WS$_2$} by magnetic proximity effect},\ }\href {https://doi.org/10.1038/s41467-019-11966-4} {\bibfield  {journal} {\bibinfo  {journal} {Nature Communications}\ }\textbf {\bibinfo {volume} {10}},\ \bibinfo {pages} {4163} (\bibinfo {year} {2019})}\BibitemShut {NoStop}%
\bibitem [{\citenamefont {Seyler}\ \emph {et~al.}(2018)\citenamefont {Seyler}, \citenamefont {Zhong}, \citenamefont {Huang}, \citenamefont {Linpeng}, \citenamefont {Wilson}, \citenamefont {Taniguchi}, \citenamefont {Watanabe}, \citenamefont {Yao}, \citenamefont {Xiao}, \citenamefont {McGuire}, \citenamefont {Fu},\ and\ \citenamefont {Xu}}]{seyler_valley_2018}%
  \BibitemOpen
  \bibfield  {author} {\bibinfo {author} {\bibfnamefont {K.~L.}\ \bibnamefont {Seyler}}, \bibinfo {author} {\bibfnamefont {D.}~\bibnamefont {Zhong}}, \bibinfo {author} {\bibfnamefont {B.}~\bibnamefont {Huang}}, \bibinfo {author} {\bibfnamefont {X.}~\bibnamefont {Linpeng}}, \bibinfo {author} {\bibfnamefont {N.~P.}\ \bibnamefont {Wilson}}, \bibinfo {author} {\bibfnamefont {T.}~\bibnamefont {Taniguchi}}, \bibinfo {author} {\bibfnamefont {K.}~\bibnamefont {Watanabe}}, \bibinfo {author} {\bibfnamefont {W.}~\bibnamefont {Yao}}, \bibinfo {author} {\bibfnamefont {D.}~\bibnamefont {Xiao}}, \bibinfo {author} {\bibfnamefont {M.~A.}\ \bibnamefont {McGuire}}, \bibinfo {author} {\bibfnamefont {K.-M.~C.}\ \bibnamefont {Fu}},\ and\ \bibinfo {author} {\bibfnamefont {X.}~\bibnamefont {Xu}},\ }\bibfield  {title} {\bibinfo {title} {Valley {Manipulation} by {Optically} {Tuning} the {Magnetic} {Proximity} {Effect} in {WSe}$_{\textrm{2}}$/{CrI}$_{\textrm{3}}$ {Heterostructures}},\ }\href
  {https://doi.org/10.1021/acs.nanolett.8b01105} {\bibfield  {journal} {\bibinfo  {journal} {Nano Letters}\ }\textbf {\bibinfo {volume} {18}},\ \bibinfo {pages} {3823} (\bibinfo {year} {2018})}\BibitemShut {NoStop}%
\bibitem [{\citenamefont {Lyons}\ \emph {et~al.}(2020)\citenamefont {Lyons}, \citenamefont {Gillard}, \citenamefont {Molina-S{\'a}nchez}, \citenamefont {Misra}, \citenamefont {Withers}, \citenamefont {Keatley}, \citenamefont {Kozikov}, \citenamefont {Taniguchi}, \citenamefont {Watanabe}, \citenamefont {Novoselov} \emph {et~al.}}]{lyons2020interplay}%
  \BibitemOpen
  \bibfield  {author} {\bibinfo {author} {\bibfnamefont {T.~P.}\ \bibnamefont {Lyons}}, \bibinfo {author} {\bibfnamefont {D.}~\bibnamefont {Gillard}}, \bibinfo {author} {\bibfnamefont {A.}~\bibnamefont {Molina-S{\'a}nchez}}, \bibinfo {author} {\bibfnamefont {A.}~\bibnamefont {Misra}}, \bibinfo {author} {\bibfnamefont {F.}~\bibnamefont {Withers}}, \bibinfo {author} {\bibfnamefont {P.~S.}\ \bibnamefont {Keatley}}, \bibinfo {author} {\bibfnamefont {A.}~\bibnamefont {Kozikov}}, \bibinfo {author} {\bibfnamefont {T.}~\bibnamefont {Taniguchi}}, \bibinfo {author} {\bibfnamefont {K.}~\bibnamefont {Watanabe}}, \bibinfo {author} {\bibfnamefont {K.~S.}\ \bibnamefont {Novoselov}}, \emph {et~al.},\ }\bibfield  {title} {\bibinfo {title} {Interplay between spin proximity effect and charge-dependent exciton dynamics in {MoSe$_2$}/{CrBr$_3$} van der waals heterostructures},\ }\href@noop {} {\bibfield  {journal} {\bibinfo  {journal} {Nature communications}\ }\textbf {\bibinfo {volume} {11}},\ \bibinfo {pages} {6021} (\bibinfo
  {year} {2020})}\BibitemShut {NoStop}%
\bibitem [{\citenamefont {Choi}\ \emph {et~al.}(2023)\citenamefont {Choi}, \citenamefont {Lane}, \citenamefont {Zhu},\ and\ \citenamefont {Crooker}}]{choi_asymmetric_2023}%
  \BibitemOpen
  \bibfield  {author} {\bibinfo {author} {\bibfnamefont {J.}~\bibnamefont {Choi}}, \bibinfo {author} {\bibfnamefont {C.}~\bibnamefont {Lane}}, \bibinfo {author} {\bibfnamefont {J.-X.}\ \bibnamefont {Zhu}},\ and\ \bibinfo {author} {\bibfnamefont {S.~A.}\ \bibnamefont {Crooker}},\ }\bibfield  {title} {\bibinfo {title} {Asymmetric magnetic proximity interactions in {MoSe$_2$}/{CrBr$_3$} van der {Waals} heterostructures},\ }\href {https://doi.org/10.1038/s41563-022-01424-w} {\bibfield  {journal} {\bibinfo  {journal} {Nature Materials}\ }\textbf {\bibinfo {volume} {22}},\ \bibinfo {pages} {305} (\bibinfo {year} {2023})}\BibitemShut {NoStop}%
\bibitem [{\citenamefont {Huang}\ \emph {et~al.}(2017)\citenamefont {Huang}, \citenamefont {Clark}, \citenamefont {Navarro-Moratalla}, \citenamefont {Klein}, \citenamefont {Cheng}, \citenamefont {Seyler}, \citenamefont {Zhong}, \citenamefont {Schmidgall}, \citenamefont {McGuire}, \citenamefont {Cobden}, \citenamefont {Yao}, \citenamefont {Xiao}, \citenamefont {Jarillo-Herrero},\ and\ \citenamefont {Xu}}]{huang_layer-dependent_2017}%
  \BibitemOpen
  \bibfield  {author} {\bibinfo {author} {\bibfnamefont {B.}~\bibnamefont {Huang}}, \bibinfo {author} {\bibfnamefont {G.}~\bibnamefont {Clark}}, \bibinfo {author} {\bibfnamefont {E.}~\bibnamefont {Navarro-Moratalla}}, \bibinfo {author} {\bibfnamefont {D.~R.}\ \bibnamefont {Klein}}, \bibinfo {author} {\bibfnamefont {R.}~\bibnamefont {Cheng}}, \bibinfo {author} {\bibfnamefont {K.~L.}\ \bibnamefont {Seyler}}, \bibinfo {author} {\bibfnamefont {D.}~\bibnamefont {Zhong}}, \bibinfo {author} {\bibfnamefont {E.}~\bibnamefont {Schmidgall}}, \bibinfo {author} {\bibfnamefont {M.~A.}\ \bibnamefont {McGuire}}, \bibinfo {author} {\bibfnamefont {D.~H.}\ \bibnamefont {Cobden}}, \bibinfo {author} {\bibfnamefont {W.}~\bibnamefont {Yao}}, \bibinfo {author} {\bibfnamefont {D.}~\bibnamefont {Xiao}}, \bibinfo {author} {\bibfnamefont {P.}~\bibnamefont {Jarillo-Herrero}},\ and\ \bibinfo {author} {\bibfnamefont {X.}~\bibnamefont {Xu}},\ }\bibfield  {title} {\bibinfo {title} {Layer-dependent ferromagnetism in a van der {Waals} crystal
  down to the monolayer limit},\ }\href {https://doi.org/10.1038/nature22391} {\bibfield  {journal} {\bibinfo  {journal} {Nature}\ }\textbf {\bibinfo {volume} {546}},\ \bibinfo {pages} {270} (\bibinfo {year} {2017})}\BibitemShut {NoStop}%
\bibitem [{\citenamefont {Gong}\ \emph {et~al.}(2017)\citenamefont {Gong}, \citenamefont {Li}, \citenamefont {Li}, \citenamefont {Ji}, \citenamefont {Stern}, \citenamefont {Xia}, \citenamefont {Cao}, \citenamefont {Bao}, \citenamefont {Wang}, \citenamefont {Wang}, \citenamefont {Qiu}, \citenamefont {Cava}, \citenamefont {Louie}, \citenamefont {Xia},\ and\ \citenamefont {Zhang}}]{gong_discovery_2017}%
  \BibitemOpen
  \bibfield  {author} {\bibinfo {author} {\bibfnamefont {C.}~\bibnamefont {Gong}}, \bibinfo {author} {\bibfnamefont {L.}~\bibnamefont {Li}}, \bibinfo {author} {\bibfnamefont {Z.}~\bibnamefont {Li}}, \bibinfo {author} {\bibfnamefont {H.}~\bibnamefont {Ji}}, \bibinfo {author} {\bibfnamefont {A.}~\bibnamefont {Stern}}, \bibinfo {author} {\bibfnamefont {Y.}~\bibnamefont {Xia}}, \bibinfo {author} {\bibfnamefont {T.}~\bibnamefont {Cao}}, \bibinfo {author} {\bibfnamefont {W.}~\bibnamefont {Bao}}, \bibinfo {author} {\bibfnamefont {C.}~\bibnamefont {Wang}}, \bibinfo {author} {\bibfnamefont {Y.}~\bibnamefont {Wang}}, \bibinfo {author} {\bibfnamefont {Z.~Q.}\ \bibnamefont {Qiu}}, \bibinfo {author} {\bibfnamefont {R.~J.}\ \bibnamefont {Cava}}, \bibinfo {author} {\bibfnamefont {S.~G.}\ \bibnamefont {Louie}}, \bibinfo {author} {\bibfnamefont {J.}~\bibnamefont {Xia}},\ and\ \bibinfo {author} {\bibfnamefont {X.}~\bibnamefont {Zhang}},\ }\bibfield  {title} {\bibinfo {title} {Discovery of intrinsic ferromagnetism in
  two-dimensional van der {Waals} crystals},\ }\href {https://doi.org/10.1038/nature22060} {\bibfield  {journal} {\bibinfo  {journal} {Nature}\ }\textbf {\bibinfo {volume} {546}},\ \bibinfo {pages} {265} (\bibinfo {year} {2017})}\BibitemShut {NoStop}%
\bibitem [{\citenamefont {Zhuo}\ \emph {et~al.}(2021)\citenamefont {Zhuo}, \citenamefont {Lei}, \citenamefont {Wu}, \citenamefont {Yu}, \citenamefont {Zhu}, \citenamefont {Cui}, \citenamefont {Sun}, \citenamefont {Ma}, \citenamefont {Shi}, \citenamefont {Wang}, \citenamefont {Wang}, \citenamefont {Wu}, \citenamefont {Ying}, \citenamefont {Wu}, \citenamefont {Wang},\ and\ \citenamefont {Chen}}]{zhuo_manipulating_2021}%
  \BibitemOpen
  \bibfield  {author} {\bibinfo {author} {\bibfnamefont {W.}~\bibnamefont {Zhuo}}, \bibinfo {author} {\bibfnamefont {B.}~\bibnamefont {Lei}}, \bibinfo {author} {\bibfnamefont {S.}~\bibnamefont {Wu}}, \bibinfo {author} {\bibfnamefont {F.}~\bibnamefont {Yu}}, \bibinfo {author} {\bibfnamefont {C.}~\bibnamefont {Zhu}}, \bibinfo {author} {\bibfnamefont {J.}~\bibnamefont {Cui}}, \bibinfo {author} {\bibfnamefont {Z.}~\bibnamefont {Sun}}, \bibinfo {author} {\bibfnamefont {D.}~\bibnamefont {Ma}}, \bibinfo {author} {\bibfnamefont {M.}~\bibnamefont {Shi}}, \bibinfo {author} {\bibfnamefont {H.}~\bibnamefont {Wang}}, \bibinfo {author} {\bibfnamefont {W.}~\bibnamefont {Wang}}, \bibinfo {author} {\bibfnamefont {T.}~\bibnamefont {Wu}}, \bibinfo {author} {\bibfnamefont {J.}~\bibnamefont {Ying}}, \bibinfo {author} {\bibfnamefont {S.}~\bibnamefont {Wu}}, \bibinfo {author} {\bibfnamefont {Z.}~\bibnamefont {Wang}},\ and\ \bibinfo {author} {\bibfnamefont {X.}~\bibnamefont {Chen}},\ }\bibfield  {title} {\bibinfo {title}
  {Manipulating {Ferromagnetism} in {Few}‐{Layered} {Cr}$_{\textrm{2}}${Ge}$_{\textrm{2}}${Te}$_{\textrm{6}}$},\ }\href {https://doi.org/10.1002/adma.202008586} {\bibfield  {journal} {\bibinfo  {journal} {Advanced Materials}\ }\textbf {\bibinfo {volume} {33}},\ \bibinfo {pages} {2008586} (\bibinfo {year} {2021})}\BibitemShut {NoStop}%
\bibitem [{\citenamefont {Tu}\ \emph {et~al.}(2022)\citenamefont {Tu}, \citenamefont {Zhou}, \citenamefont {Ersevim}, \citenamefont {Arachchige}, \citenamefont {Hanbicki}, \citenamefont {Friedman}, \citenamefont {Mandrus}, \citenamefont {Ouyang}, \citenamefont {Žutić},\ and\ \citenamefont {Gong}}]{tu_spinorbit_2022}%
  \BibitemOpen
  \bibfield  {author} {\bibinfo {author} {\bibfnamefont {Z.}~\bibnamefont {Tu}}, \bibinfo {author} {\bibfnamefont {T.}~\bibnamefont {Zhou}}, \bibinfo {author} {\bibfnamefont {T.}~\bibnamefont {Ersevim}}, \bibinfo {author} {\bibfnamefont {H.~S.}\ \bibnamefont {Arachchige}}, \bibinfo {author} {\bibfnamefont {A.~T.}\ \bibnamefont {Hanbicki}}, \bibinfo {author} {\bibfnamefont {A.~L.}\ \bibnamefont {Friedman}}, \bibinfo {author} {\bibfnamefont {D.}~\bibnamefont {Mandrus}}, \bibinfo {author} {\bibfnamefont {M.}~\bibnamefont {Ouyang}}, \bibinfo {author} {\bibfnamefont {I.}~\bibnamefont {Žutić}},\ and\ \bibinfo {author} {\bibfnamefont {C.}~\bibnamefont {Gong}},\ }\bibfield  {title} {\bibinfo {title} {Spin–orbit coupling proximity effect in {MoS$_2$}/{Fe$_3$GeTe$_2$} heterostructures},\ }\href {https://doi.org/10.1063/5.0080505} {\bibfield  {journal} {\bibinfo  {journal} {Applied Physics Letters}\ }\textbf {\bibinfo {volume} {120}},\ \bibinfo {pages} {043102} (\bibinfo {year} {2022})}\BibitemShut {NoStop}%
\bibitem [{\citenamefont {Gupta}\ \emph {et~al.}(2020)\citenamefont {Gupta}, \citenamefont {Cham}, \citenamefont {Stiehl}, \citenamefont {Bose}, \citenamefont {Mittelstaedt}, \citenamefont {Kang}, \citenamefont {Jiang}, \citenamefont {Mak}, \citenamefont {Shan}, \citenamefont {Buhrman},\ and\ \citenamefont {Ralph}}]{gupta_manipulation_2020}%
  \BibitemOpen
  \bibfield  {author} {\bibinfo {author} {\bibfnamefont {V.}~\bibnamefont {Gupta}}, \bibinfo {author} {\bibfnamefont {T.~M.}\ \bibnamefont {Cham}}, \bibinfo {author} {\bibfnamefont {G.~M.}\ \bibnamefont {Stiehl}}, \bibinfo {author} {\bibfnamefont {A.}~\bibnamefont {Bose}}, \bibinfo {author} {\bibfnamefont {J.~A.}\ \bibnamefont {Mittelstaedt}}, \bibinfo {author} {\bibfnamefont {K.}~\bibnamefont {Kang}}, \bibinfo {author} {\bibfnamefont {S.}~\bibnamefont {Jiang}}, \bibinfo {author} {\bibfnamefont {K.~F.}\ \bibnamefont {Mak}}, \bibinfo {author} {\bibfnamefont {J.}~\bibnamefont {Shan}}, \bibinfo {author} {\bibfnamefont {R.~A.}\ \bibnamefont {Buhrman}},\ and\ \bibinfo {author} {\bibfnamefont {D.~C.}\ \bibnamefont {Ralph}},\ }\bibfield  {title} {\bibinfo {title} {Manipulation of the van der {Waals} {Magnet} {Cr}$_{\textrm{2}}${Ge}$_{\textrm{2}}${Te}$_{\textrm{6}}$ by {Spin}–{Orbit} {Torques}},\ }\href {https://doi.org/10.1021/acs.nanolett.0c02965} {\bibfield  {journal} {\bibinfo  {journal} {Nano Letters}\
  }\textbf {\bibinfo {volume} {20}},\ \bibinfo {pages} {7482} (\bibinfo {year} {2020})}\BibitemShut {NoStop}%
\bibitem [{\citenamefont {Goff}\ \emph {et~al.}(2024)\citenamefont {Goff}, \citenamefont {Zhou}, \citenamefont {Bishop}, \citenamefont {Bailey-Crandell}, \citenamefont {Robinson}, \citenamefont {Kawakami},\ and\ \citenamefont {Gupta}}]{goff_scanning_2024}%
  \BibitemOpen
  \bibfield  {author} {\bibinfo {author} {\bibfnamefont {B.~M.}\ \bibnamefont {Goff}}, \bibinfo {author} {\bibfnamefont {W.}~\bibnamefont {Zhou}}, \bibinfo {author} {\bibfnamefont {A.~J.}\ \bibnamefont {Bishop}}, \bibinfo {author} {\bibfnamefont {R.}~\bibnamefont {Bailey-Crandell}}, \bibinfo {author} {\bibfnamefont {K.}~\bibnamefont {Robinson}}, \bibinfo {author} {\bibfnamefont {R.~K.}\ \bibnamefont {Kawakami}},\ and\ \bibinfo {author} {\bibfnamefont {J.~A.}\ \bibnamefont {Gupta}},\ }\bibfield  {title} {\bibinfo {title} {Scanning tunneling microscopy study of epitaxial {Fe}$_{\textrm{3}}${GeTe}$_{\textrm{2}}$ monolayers on {Bi}$_{\textrm{2}}${Te}$_{\textrm{3}}$},\ }\href {https://doi.org/10.1088/2053-1583/ad1c6d} {\bibfield  {journal} {\bibinfo  {journal} {2D Materials}\ }\textbf {\bibinfo {volume} {11}},\ \bibinfo {pages} {025012} (\bibinfo {year} {2024})}\BibitemShut {NoStop}%
\bibitem [{\citenamefont {Cham}\ \emph {et~al.}(2025)\citenamefont {Cham}, \citenamefont {Chica}, \citenamefont {Huang}, \citenamefont {Watanabe}, \citenamefont {Taniguchi}, \citenamefont {Roy}, \citenamefont {Luo},\ and\ \citenamefont {Ralph}}]{cham_spin-filter_2025}%
  \BibitemOpen
  \bibfield  {author} {\bibinfo {author} {\bibfnamefont {T.~M.~J.}\ \bibnamefont {Cham}}, \bibinfo {author} {\bibfnamefont {D.~G.}\ \bibnamefont {Chica}}, \bibinfo {author} {\bibfnamefont {X.}~\bibnamefont {Huang}}, \bibinfo {author} {\bibfnamefont {K.}~\bibnamefont {Watanabe}}, \bibinfo {author} {\bibfnamefont {T.}~\bibnamefont {Taniguchi}}, \bibinfo {author} {\bibfnamefont {X.}~\bibnamefont {Roy}}, \bibinfo {author} {\bibfnamefont {Y.~K.}\ \bibnamefont {Luo}},\ and\ \bibinfo {author} {\bibfnamefont {D.~C.}\ \bibnamefont {Ralph}},\ }\bibfield  {title} {\bibinfo {title} {Spin-filter tunneling detection of antiferromagnetic resonance with electrically tunable damping},\ }\href {https://doi.org/10.1126/science.adq8590} {\bibfield  {journal} {\bibinfo  {journal} {Science}\ }\textbf {\bibinfo {volume} {389}},\ \bibinfo {pages} {479} (\bibinfo {year} {2025})}\BibitemShut {NoStop}%
\bibitem [{\citenamefont {Gibertini}\ \emph {et~al.}(2019)\citenamefont {Gibertini}, \citenamefont {Koperski}, \citenamefont {Morpurgo},\ and\ \citenamefont {Novoselov}}]{gibertini_magnetic_2019}%
  \BibitemOpen
  \bibfield  {author} {\bibinfo {author} {\bibfnamefont {M.}~\bibnamefont {Gibertini}}, \bibinfo {author} {\bibfnamefont {M.}~\bibnamefont {Koperski}}, \bibinfo {author} {\bibfnamefont {A.~F.}\ \bibnamefont {Morpurgo}},\ and\ \bibinfo {author} {\bibfnamefont {K.~S.}\ \bibnamefont {Novoselov}},\ }\bibfield  {title} {\bibinfo {title} {Magnetic 2{D} materials and heterostructures},\ }\href {https://doi.org/10.1038/s41565-019-0438-6} {\bibfield  {journal} {\bibinfo  {journal} {Nature Nanotechnology}\ }\textbf {\bibinfo {volume} {14}},\ \bibinfo {pages} {408} (\bibinfo {year} {2019})}\BibitemShut {NoStop}%
\bibitem [{\citenamefont {Sun}\ \emph {et~al.}(2021)\citenamefont {Sun}, \citenamefont {Zhou}, \citenamefont {Jiang}, \citenamefont {Li}, \citenamefont {Qiu}, \citenamefont {Xiao}, \citenamefont {Liu}, \citenamefont {Ma}, \citenamefont {Luo}, \citenamefont {Sun},\ and\ \citenamefont {Sheng}}]{sun_ultra-long_2021}%
  \BibitemOpen
  \bibfield  {author} {\bibinfo {author} {\bibfnamefont {T.}~\bibnamefont {Sun}}, \bibinfo {author} {\bibfnamefont {C.}~\bibnamefont {Zhou}}, \bibinfo {author} {\bibfnamefont {Z.}~\bibnamefont {Jiang}}, \bibinfo {author} {\bibfnamefont {X.}~\bibnamefont {Li}}, \bibinfo {author} {\bibfnamefont {K.}~\bibnamefont {Qiu}}, \bibinfo {author} {\bibfnamefont {R.}~\bibnamefont {Xiao}}, \bibinfo {author} {\bibfnamefont {C.}~\bibnamefont {Liu}}, \bibinfo {author} {\bibfnamefont {Z.}~\bibnamefont {Ma}}, \bibinfo {author} {\bibfnamefont {X.}~\bibnamefont {Luo}}, \bibinfo {author} {\bibfnamefont {Y.}~\bibnamefont {Sun}},\ and\ \bibinfo {author} {\bibfnamefont {Z.}~\bibnamefont {Sheng}},\ }\bibfield  {title} {\bibinfo {title} {Ultra-long spin relaxation in two-dimensional ferromagnet {Cr}$_{\textrm{2}}${Ge}$_{\textrm{2}}${Te}$_{\textrm{6}}$ flake},\ }\href {https://doi.org/10.1088/2053-1583/ac2ab3} {\bibfield  {journal} {\bibinfo  {journal} {2D Materials}\ }\textbf {\bibinfo {volume} {8}},\ \bibinfo {pages} {045040}
  (\bibinfo {year} {2021})}\BibitemShut {NoStop}%
\bibitem [{\citenamefont {Sutcliffe}\ \emph {et~al.}(2023)\citenamefont {Sutcliffe}, \citenamefont {Sun}, \citenamefont {Verzhbitskiy}, \citenamefont {Griepe}, \citenamefont {Atxitia}, \citenamefont {Eda}, \citenamefont {Santos},\ and\ \citenamefont {Johansson}}]{sutcliffe_transient_2023}%
  \BibitemOpen
  \bibfield  {author} {\bibinfo {author} {\bibfnamefont {E.}~\bibnamefont {Sutcliffe}}, \bibinfo {author} {\bibfnamefont {X.}~\bibnamefont {Sun}}, \bibinfo {author} {\bibfnamefont {I.}~\bibnamefont {Verzhbitskiy}}, \bibinfo {author} {\bibfnamefont {T.}~\bibnamefont {Griepe}}, \bibinfo {author} {\bibfnamefont {U.}~\bibnamefont {Atxitia}}, \bibinfo {author} {\bibfnamefont {G.}~\bibnamefont {Eda}}, \bibinfo {author} {\bibfnamefont {E.~J.~G.}\ \bibnamefont {Santos}},\ and\ \bibinfo {author} {\bibfnamefont {J.~O.}\ \bibnamefont {Johansson}},\ }\bibfield  {title} {\bibinfo {title} {Transient magneto-optical spectrum of photoexcited electrons in the van der {W}aals ferromagnet {Cr}$_{\textrm{2}}${Ge}$_{\textrm{2}}${Te}$_{\textrm{6}}$},\ }\href {https://doi.org/10.1103/PhysRevB.107.174432} {\bibfield  {journal} {\bibinfo  {journal} {Physical Review B}\ }\textbf {\bibinfo {volume} {107}},\ \bibinfo {pages} {174432} (\bibinfo {year} {2023})}\BibitemShut {NoStop}%
\bibitem [{\citenamefont {Dabrowski}\ \emph {et~al.}(2025)\citenamefont {Dabrowski}, \citenamefont {Haldar}, \citenamefont {Khan}, \citenamefont {Keatley}, \citenamefont {Sagkovits}, \citenamefont {Xue}, \citenamefont {Freeman}, \citenamefont {Verzhbitskiy}, \citenamefont {Griepe}, \citenamefont {Atxitia}, \citenamefont {Eda}, \citenamefont {Kurebayashi}, \citenamefont {Santos},\ and\ \citenamefont {Hicken}}]{dabrowski_ultrafast_2025}%
  \BibitemOpen
  \bibfield  {author} {\bibinfo {author} {\bibfnamefont {M.}~\bibnamefont {Dabrowski}}, \bibinfo {author} {\bibfnamefont {S.}~\bibnamefont {Haldar}}, \bibinfo {author} {\bibfnamefont {S.}~\bibnamefont {Khan}}, \bibinfo {author} {\bibfnamefont {P.~S.}\ \bibnamefont {Keatley}}, \bibinfo {author} {\bibfnamefont {D.}~\bibnamefont {Sagkovits}}, \bibinfo {author} {\bibfnamefont {Z.}~\bibnamefont {Xue}}, \bibinfo {author} {\bibfnamefont {C.}~\bibnamefont {Freeman}}, \bibinfo {author} {\bibfnamefont {I.}~\bibnamefont {Verzhbitskiy}}, \bibinfo {author} {\bibfnamefont {T.}~\bibnamefont {Griepe}}, \bibinfo {author} {\bibfnamefont {U.}~\bibnamefont {Atxitia}}, \bibinfo {author} {\bibfnamefont {G.}~\bibnamefont {Eda}}, \bibinfo {author} {\bibfnamefont {H.}~\bibnamefont {Kurebayashi}}, \bibinfo {author} {\bibfnamefont {E.~J.~G.}\ \bibnamefont {Santos}},\ and\ \bibinfo {author} {\bibfnamefont {R.~J.}\ \bibnamefont {Hicken}},\ }\bibfield  {title} {\bibinfo {title} {Ultrafast thermo-optical control of spins in a 2{D} van der
  {W}aals semiconductor},\ }\href {https://doi.org/10.1038/s41467-025-58065-1} {\bibfield  {journal} {\bibinfo  {journal} {Nature Communications}\ }\textbf {\bibinfo {volume} {16}},\ \bibinfo {pages} {2797} (\bibinfo {year} {2025})}\BibitemShut {NoStop}%
\bibitem [{\citenamefont {Verzhbitskiy}\ \emph {et~al.}(2020)\citenamefont {Verzhbitskiy}, \citenamefont {Kurebayashi}, \citenamefont {Cheng}, \citenamefont {Zhou}, \citenamefont {Khan}, \citenamefont {Feng},\ and\ \citenamefont {Eda}}]{verzhbitskiy_controlling_2020}%
  \BibitemOpen
  \bibfield  {author} {\bibinfo {author} {\bibfnamefont {I.~A.}\ \bibnamefont {Verzhbitskiy}}, \bibinfo {author} {\bibfnamefont {H.}~\bibnamefont {Kurebayashi}}, \bibinfo {author} {\bibfnamefont {H.}~\bibnamefont {Cheng}}, \bibinfo {author} {\bibfnamefont {J.}~\bibnamefont {Zhou}}, \bibinfo {author} {\bibfnamefont {S.}~\bibnamefont {Khan}}, \bibinfo {author} {\bibfnamefont {Y.~P.}\ \bibnamefont {Feng}},\ and\ \bibinfo {author} {\bibfnamefont {G.}~\bibnamefont {Eda}},\ }\bibfield  {title} {\bibinfo {title} {Controlling the magnetic anisotropy in {Cr$_2$Ge$_2$Te$_6$} by electrostatic gating},\ }\href {https://doi.org/10.1038/s41928-020-0427-7} {\bibfield  {journal} {\bibinfo  {journal} {Nature Electronics}\ }\textbf {\bibinfo {volume} {3}},\ \bibinfo {pages} {460} (\bibinfo {year} {2020})}\BibitemShut {NoStop}%
\bibitem [{Sup()}]{SupplMat}%
  \BibitemOpen
  \href@noop {} {}\bibinfo {note} {See Supplemental Material for discussion of the heterostructure fabrication, photoluminescence measurements, detailed temperature-dependent MOKE loops, TRKE measurement details and fitting formula, TRKE data at different temperatures, LLG equation modelling, DFT calculations, and photocurrent measurements, which includes Refs.~ \cite{hairer_solving_1996,virtanen_scipy_2020,coleman_interior_1996,furthmuller1996dimer,kresse1996efficiency,perdew1996generalized,blochl1994projector,heyd2003hybrid}.}\BibitemShut {Stop}%
\bibitem [{\citenamefont {Rahman}\ \emph {et~al.}(2021)\citenamefont {Rahman}, \citenamefont {Liu}, \citenamefont {Wang}, \citenamefont {Tang},\ and\ \citenamefont {Lu}}]{rahman_giant_2021}%
  \BibitemOpen
  \bibfield  {author} {\bibinfo {author} {\bibfnamefont {S.}~\bibnamefont {Rahman}}, \bibinfo {author} {\bibfnamefont {B.}~\bibnamefont {Liu}}, \bibinfo {author} {\bibfnamefont {B.}~\bibnamefont {Wang}}, \bibinfo {author} {\bibfnamefont {Y.}~\bibnamefont {Tang}},\ and\ \bibinfo {author} {\bibfnamefont {Y.}~\bibnamefont {Lu}},\ }\bibfield  {title} {\bibinfo {title} {Giant {Photoluminescence} {Enhancement} and {Resonant} {Charge} {Transfer} in {Atomically} {Thin} {Two}-{Dimensional} {Cr}$_{\textrm{2}}${Ge}$_{\textrm{2}}${Te}$_{\textrm{6}}$/{WS}$_{\textrm{2}}$ {Heterostructures}},\ }\href {https://doi.org/10.1021/acsami.0c20110} {\bibfield  {journal} {\bibinfo  {journal} {ACS Applied Materials \& Interfaces}\ }\textbf {\bibinfo {volume} {13}},\ \bibinfo {pages} {7423} (\bibinfo {year} {2021})}\BibitemShut {NoStop}%
\bibitem [{\citenamefont {Zhang}\ \emph {et~al.}(2022)\citenamefont {Zhang}, \citenamefont {Zhao}, \citenamefont {Wang}, \citenamefont {Xiong}, \citenamefont {Liu}, \citenamefont {Xi}, \citenamefont {Li}, \citenamefont {Lei}, \citenamefont {Han},\ and\ \citenamefont {Wang}}]{zhang_electrically_2022}%
  \BibitemOpen
  \bibfield  {author} {\bibinfo {author} {\bibfnamefont {T.}~\bibnamefont {Zhang}}, \bibinfo {author} {\bibfnamefont {S.}~\bibnamefont {Zhao}}, \bibinfo {author} {\bibfnamefont {A.}~\bibnamefont {Wang}}, \bibinfo {author} {\bibfnamefont {Z.}~\bibnamefont {Xiong}}, \bibinfo {author} {\bibfnamefont {Y.}~\bibnamefont {Liu}}, \bibinfo {author} {\bibfnamefont {M.}~\bibnamefont {Xi}}, \bibinfo {author} {\bibfnamefont {S.}~\bibnamefont {Li}}, \bibinfo {author} {\bibfnamefont {H.}~\bibnamefont {Lei}}, \bibinfo {author} {\bibfnamefont {Z.~V.}\ \bibnamefont {Han}},\ and\ \bibinfo {author} {\bibfnamefont {F.}~\bibnamefont {Wang}},\ }\bibfield  {title} {\bibinfo {title} {Electrically and {Magnetically} {Tunable} {Valley} {Polarization} in {Monolayer} {MoSe}$_{\textrm{2}}$ {Proximitized} by a {2D} {Ferromagnetic} {Semiconductor}},\ }\href {https://doi.org/10.1002/adfm.202204779} {\bibfield  {journal} {\bibinfo  {journal} {Advanced Functional Materials}\ }\textbf {\bibinfo {volume} {32}},\ \bibinfo {pages} {2204779}
  (\bibinfo {year} {2022})}\BibitemShut {NoStop}%
\bibitem [{\citenamefont {Weisheit}\ \emph {et~al.}(2007)\citenamefont {Weisheit}, \citenamefont {F{\"a}hler}, \citenamefont {Marty}, \citenamefont {Souche}, \citenamefont {Poinsignon},\ and\ \citenamefont {Givord}}]{weisheit_electric_2007}%
  \BibitemOpen
  \bibfield  {author} {\bibinfo {author} {\bibfnamefont {M.}~\bibnamefont {Weisheit}}, \bibinfo {author} {\bibfnamefont {S.}~\bibnamefont {F{\"a}hler}}, \bibinfo {author} {\bibfnamefont {A.}~\bibnamefont {Marty}}, \bibinfo {author} {\bibfnamefont {Y.}~\bibnamefont {Souche}}, \bibinfo {author} {\bibfnamefont {C.}~\bibnamefont {Poinsignon}},\ and\ \bibinfo {author} {\bibfnamefont {D.}~\bibnamefont {Givord}},\ }\bibfield  {title} {\bibinfo {title} {Electric {Field}-{Induced} {Modification} of {Magnetism} in {Thin}-{Film} {Ferromagnets}},\ }\href {https://doi.org/10.1126/science.1136629} {\bibfield  {journal} {\bibinfo  {journal} {Science}\ }\textbf {\bibinfo {volume} {315}},\ \bibinfo {pages} {349} (\bibinfo {year} {2007})}\BibitemShut {NoStop}%
\bibitem [{\citenamefont {Duan}\ \emph {et~al.}(2008)\citenamefont {Duan}, \citenamefont {Velev}, \citenamefont {Sabirianov}, \citenamefont {Zhu}, \citenamefont {Chu}, \citenamefont {Jaswal},\ and\ \citenamefont {Tsymbal}}]{duan_surface_2008}%
  \BibitemOpen
  \bibfield  {author} {\bibinfo {author} {\bibfnamefont {C.-G.}\ \bibnamefont {Duan}}, \bibinfo {author} {\bibfnamefont {J.~P.}\ \bibnamefont {Velev}}, \bibinfo {author} {\bibfnamefont {R.~F.}\ \bibnamefont {Sabirianov}}, \bibinfo {author} {\bibfnamefont {Z.}~\bibnamefont {Zhu}}, \bibinfo {author} {\bibfnamefont {J.}~\bibnamefont {Chu}}, \bibinfo {author} {\bibfnamefont {S.~S.}\ \bibnamefont {Jaswal}},\ and\ \bibinfo {author} {\bibfnamefont {E.~Y.}\ \bibnamefont {Tsymbal}},\ }\bibfield  {title} {\bibinfo {title} {Surface {Magnetoelectric} {Effect} in {Ferromagnetic} {Metal} {Films}},\ }\href {https://doi.org/10.1103/PhysRevLett.101.137201} {\bibfield  {journal} {\bibinfo  {journal} {Physical Review Letters}\ }\textbf {\bibinfo {volume} {101}},\ \bibinfo {pages} {137201} (\bibinfo {year} {2008})}\BibitemShut {NoStop}%
\bibitem [{\citenamefont {Nakamura}\ \emph {et~al.}(2009)\citenamefont {Nakamura}, \citenamefont {Shimabukuro}, \citenamefont {Fujiwara}, \citenamefont {Akiyama}, \citenamefont {Ito},\ and\ \citenamefont {Freeman}}]{nakamura_giant_2009}%
  \BibitemOpen
  \bibfield  {author} {\bibinfo {author} {\bibfnamefont {K.}~\bibnamefont {Nakamura}}, \bibinfo {author} {\bibfnamefont {R.}~\bibnamefont {Shimabukuro}}, \bibinfo {author} {\bibfnamefont {Y.}~\bibnamefont {Fujiwara}}, \bibinfo {author} {\bibfnamefont {T.}~\bibnamefont {Akiyama}}, \bibinfo {author} {\bibfnamefont {T.}~\bibnamefont {Ito}},\ and\ \bibinfo {author} {\bibfnamefont {A.~J.}\ \bibnamefont {Freeman}},\ }\bibfield  {title} {\bibinfo {title} {Giant {Modification} of the {Magnetocrystalline} {Anisotropy} in {Transition}-{Metal} {Monolayers} by an {External} {Electric} {Field}},\ }\href {https://doi.org/10.1103/PhysRevLett.102.187201} {\bibfield  {journal} {\bibinfo  {journal} {Physical Review Letters}\ }\textbf {\bibinfo {volume} {102}},\ \bibinfo {pages} {187201} (\bibinfo {year} {2009})}\BibitemShut {NoStop}%
\bibitem [{\citenamefont {Jiang}\ \emph {et~al.}(2018)\citenamefont {Jiang}, \citenamefont {Li}, \citenamefont {Wang}, \citenamefont {Mak},\ and\ \citenamefont {Shan}}]{jiang_electrical_2018}%
  \BibitemOpen
  \bibfield  {author} {\bibinfo {author} {\bibfnamefont {S.}~\bibnamefont {Jiang}}, \bibinfo {author} {\bibfnamefont {L.}~\bibnamefont {Li}}, \bibinfo {author} {\bibfnamefont {Z.}~\bibnamefont {Wang}}, \bibinfo {author} {\bibfnamefont {K.~F.}\ \bibnamefont {Mak}},\ and\ \bibinfo {author} {\bibfnamefont {J.}~\bibnamefont {Shan}},\ }\bibfield  {title} {\bibinfo {title} {Electrical control of 2{D} magnetism in bilayer {CrI}$_{\textrm{3}}$},\ }\href {https://doi.org/10.1038/s41565-018-0121-3} {\bibfield  {journal} {\bibinfo  {journal} {Nature Nanotechnology}\ }\textbf {\bibinfo {volume} {13}},\ \bibinfo {pages} {549} (\bibinfo {year} {2018})}\BibitemShut {NoStop}%
\bibitem [{\citenamefont {Xie}\ \emph {et~al.}(2024)\citenamefont {Xie}, \citenamefont {Zhang}, \citenamefont {Zhang}, \citenamefont {Nagarajan}, \citenamefont {Zhao}, \citenamefont {Kim}, \citenamefont {Sanborn}, \citenamefont {Qi}, \citenamefont {Chen}, \citenamefont {Kahn}, \citenamefont {Watanabe}, \citenamefont {Taniguchi}, \citenamefont {Zettl}, \citenamefont {Crommie}, \citenamefont {Analytis},\ and\ \citenamefont {Wang}}]{xie_low_2024}%
  \BibitemOpen
  \bibfield  {author} {\bibinfo {author} {\bibfnamefont {J.}~\bibnamefont {Xie}}, \bibinfo {author} {\bibfnamefont {Z.}~\bibnamefont {Zhang}}, \bibinfo {author} {\bibfnamefont {H.}~\bibnamefont {Zhang}}, \bibinfo {author} {\bibfnamefont {V.}~\bibnamefont {Nagarajan}}, \bibinfo {author} {\bibfnamefont {W.}~\bibnamefont {Zhao}}, \bibinfo {author} {\bibfnamefont {H.-L.}\ \bibnamefont {Kim}}, \bibinfo {author} {\bibfnamefont {C.}~\bibnamefont {Sanborn}}, \bibinfo {author} {\bibfnamefont {R.}~\bibnamefont {Qi}}, \bibinfo {author} {\bibfnamefont {S.}~\bibnamefont {Chen}}, \bibinfo {author} {\bibfnamefont {S.}~\bibnamefont {Kahn}}, \bibinfo {author} {\bibfnamefont {K.}~\bibnamefont {Watanabe}}, \bibinfo {author} {\bibfnamefont {T.}~\bibnamefont {Taniguchi}}, \bibinfo {author} {\bibfnamefont {A.}~\bibnamefont {Zettl}}, \bibinfo {author} {\bibfnamefont {M.~F.}\ \bibnamefont {Crommie}}, \bibinfo {author} {\bibfnamefont {J.}~\bibnamefont {Analytis}},\ and\ \bibinfo {author} {\bibfnamefont {F.}~\bibnamefont {Wang}},\
  }\bibfield  {title} {\bibinfo {title} {Low {Resistance} {Contact} to {P}-{Type} {Monolayer} {WSe}$_{\textrm{2}}$},\ }\href {https://doi.org/10.1021/acs.nanolett.3c04195} {\bibfield  {journal} {\bibinfo  {journal} {Nano Letters}\ }\textbf {\bibinfo {volume} {24}},\ \bibinfo {pages} {5937} (\bibinfo {year} {2024})}\BibitemShut {NoStop}%
\bibitem [{\citenamefont {Sternbach}\ \emph {et~al.}(2023)\citenamefont {Sternbach}, \citenamefont {Vitalone}, \citenamefont {Shabani}, \citenamefont {Zhang}, \citenamefont {Darlington}, \citenamefont {Moore}, \citenamefont {Chae}, \citenamefont {Seewald}, \citenamefont {Xu}, \citenamefont {Dean}, \citenamefont {Zhu}, \citenamefont {Rubio}, \citenamefont {Hone}, \citenamefont {Pasupathy}, \citenamefont {Schuck},\ and\ \citenamefont {Basov}}]{sternbach_quenched_2023}%
  \BibitemOpen
  \bibfield  {author} {\bibinfo {author} {\bibfnamefont {A.~J.}\ \bibnamefont {Sternbach}}, \bibinfo {author} {\bibfnamefont {R.~A.}\ \bibnamefont {Vitalone}}, \bibinfo {author} {\bibfnamefont {S.}~\bibnamefont {Shabani}}, \bibinfo {author} {\bibfnamefont {J.}~\bibnamefont {Zhang}}, \bibinfo {author} {\bibfnamefont {T.~P.}\ \bibnamefont {Darlington}}, \bibinfo {author} {\bibfnamefont {S.~L.}\ \bibnamefont {Moore}}, \bibinfo {author} {\bibfnamefont {S.~H.}\ \bibnamefont {Chae}}, \bibinfo {author} {\bibfnamefont {E.}~\bibnamefont {Seewald}}, \bibinfo {author} {\bibfnamefont {X.}~\bibnamefont {Xu}}, \bibinfo {author} {\bibfnamefont {C.~R.}\ \bibnamefont {Dean}}, \bibinfo {author} {\bibfnamefont {X.}~\bibnamefont {Zhu}}, \bibinfo {author} {\bibfnamefont {A.}~\bibnamefont {Rubio}}, \bibinfo {author} {\bibfnamefont {J.}~\bibnamefont {Hone}}, \bibinfo {author} {\bibfnamefont {A.~N.}\ \bibnamefont {Pasupathy}}, \bibinfo {author} {\bibfnamefont {P.~J.}\ \bibnamefont {Schuck}},\ and\ \bibinfo {author} {\bibfnamefont
  {D.~N.}\ \bibnamefont {Basov}},\ }\bibfield  {title} {\bibinfo {title} {Quenched {Excitons} in {WSe}$_{\textrm{2}}$/$\alpha$-{RuCl}$_{\textrm{3}}$ {Heterostructures} {Revealed} by {Multimessenger} {Nanoscopy}},\ }\href {https://doi.org/10.1021/acs.nanolett.3c00974} {\bibfield  {journal} {\bibinfo  {journal} {Nano Letters}\ }\textbf {\bibinfo {volume} {23}},\ \bibinfo {pages} {5070} (\bibinfo {year} {2023})}\BibitemShut {NoStop}%
\bibitem [{\citenamefont {Hairer}\ and\ \citenamefont {Wanner}(1996)}]{hairer_solving_1996}%
  \BibitemOpen
  \bibfield  {author} {\bibinfo {author} {\bibfnamefont {E.}~\bibnamefont {Hairer}}\ and\ \bibinfo {author} {\bibfnamefont {G.}~\bibnamefont {Wanner}},\ }\href {https://doi.org/10.1007/978-3-642-05221-7} {\emph {\bibinfo {title} {Solving {Ordinary} {Differential} {Equations} {II}}}},\ \bibinfo {series} {Springer {Series} in {Computational} {Mathematics}}, Vol.~\bibinfo {volume} {14}\ (\bibinfo  {publisher} {Springer},\ \bibinfo {address} {Berlin, Heidelberg},\ \bibinfo {year} {1996})\BibitemShut {NoStop}%
\bibitem [{\citenamefont {Virtanen}\ \emph {et~al.}(2020)\citenamefont {Virtanen}, \citenamefont {Gommers}, \citenamefont {Oliphant}, \citenamefont {Haberland}, \citenamefont {Reddy}, \citenamefont {Cournapeau}, \citenamefont {Burovski}, \citenamefont {Peterson}, \citenamefont {Weckesser}, \citenamefont {Bright}, \citenamefont {van~der Walt}, \citenamefont {Brett}, \citenamefont {Wilson}, \citenamefont {Millman}, \citenamefont {Mayorov}, \citenamefont {Nelson}, \citenamefont {Jones}, \citenamefont {Kern}, \citenamefont {Larson}, \citenamefont {Carey}, \citenamefont {Polat}, \citenamefont {Feng}, \citenamefont {Moore}, \citenamefont {VanderPlas}, \citenamefont {Laxalde}, \citenamefont {Perktold}, \citenamefont {Cimrman}, \citenamefont {Henriksen}, \citenamefont {Quintero}, \citenamefont {Harris}, \citenamefont {Archibald}, \citenamefont {Ribeiro}, \citenamefont {Pedregosa},\ and\ \citenamefont {van Mulbregt}}]{virtanen_scipy_2020}%
  \BibitemOpen
  \bibfield  {author} {\bibinfo {author} {\bibfnamefont {P.}~\bibnamefont {Virtanen}}, \bibinfo {author} {\bibfnamefont {R.}~\bibnamefont {Gommers}}, \bibinfo {author} {\bibfnamefont {T.~E.}\ \bibnamefont {Oliphant}}, \bibinfo {author} {\bibfnamefont {M.}~\bibnamefont {Haberland}}, \bibinfo {author} {\bibfnamefont {T.}~\bibnamefont {Reddy}}, \bibinfo {author} {\bibfnamefont {D.}~\bibnamefont {Cournapeau}}, \bibinfo {author} {\bibfnamefont {E.}~\bibnamefont {Burovski}}, \bibinfo {author} {\bibfnamefont {P.}~\bibnamefont {Peterson}}, \bibinfo {author} {\bibfnamefont {W.}~\bibnamefont {Weckesser}}, \bibinfo {author} {\bibfnamefont {J.}~\bibnamefont {Bright}}, \bibinfo {author} {\bibfnamefont {S.~J.}\ \bibnamefont {van~der Walt}}, \bibinfo {author} {\bibfnamefont {M.}~\bibnamefont {Brett}}, \bibinfo {author} {\bibfnamefont {J.}~\bibnamefont {Wilson}}, \bibinfo {author} {\bibfnamefont {K.~J.}\ \bibnamefont {Millman}}, \bibinfo {author} {\bibfnamefont {N.}~\bibnamefont {Mayorov}}, \bibinfo {author} {\bibfnamefont
  {A.~R.~J.}\ \bibnamefont {Nelson}}, \bibinfo {author} {\bibfnamefont {E.}~\bibnamefont {Jones}}, \bibinfo {author} {\bibfnamefont {R.}~\bibnamefont {Kern}}, \bibinfo {author} {\bibfnamefont {E.}~\bibnamefont {Larson}}, \bibinfo {author} {\bibfnamefont {C.~J.}\ \bibnamefont {Carey}}, \bibinfo {author} {\bibfnamefont {I.}~\bibnamefont {Polat}}, \bibinfo {author} {\bibfnamefont {Y.}~\bibnamefont {Feng}}, \bibinfo {author} {\bibfnamefont {E.~W.}\ \bibnamefont {Moore}}, \bibinfo {author} {\bibfnamefont {J.}~\bibnamefont {VanderPlas}}, \bibinfo {author} {\bibfnamefont {D.}~\bibnamefont {Laxalde}}, \bibinfo {author} {\bibfnamefont {J.}~\bibnamefont {Perktold}}, \bibinfo {author} {\bibfnamefont {R.}~\bibnamefont {Cimrman}}, \bibinfo {author} {\bibfnamefont {I.}~\bibnamefont {Henriksen}}, \bibinfo {author} {\bibfnamefont {E.~A.}\ \bibnamefont {Quintero}}, \bibinfo {author} {\bibfnamefont {C.~R.}\ \bibnamefont {Harris}}, \bibinfo {author} {\bibfnamefont {A.~M.}\ \bibnamefont {Archibald}}, \bibinfo {author}
  {\bibfnamefont {A.~H.}\ \bibnamefont {Ribeiro}}, \bibinfo {author} {\bibfnamefont {F.}~\bibnamefont {Pedregosa}},\ and\ \bibinfo {author} {\bibfnamefont {P.}~\bibnamefont {van Mulbregt}},\ }\bibfield  {title} {\bibinfo {title} {{SciPy} 1.0: fundamental algorithms for scientific computing in {Python}},\ }\href {https://doi.org/10.1038/s41592-019-0686-2} {\bibfield  {journal} {\bibinfo  {journal} {Nature Methods}\ }\textbf {\bibinfo {volume} {17}},\ \bibinfo {pages} {261} (\bibinfo {year} {2020})}\BibitemShut {NoStop}%
\bibitem [{\citenamefont {Coleman}\ and\ \citenamefont {Li}(1996)}]{coleman_interior_1996}%
  \BibitemOpen
  \bibfield  {author} {\bibinfo {author} {\bibfnamefont {T.~F.}\ \bibnamefont {Coleman}}\ and\ \bibinfo {author} {\bibfnamefont {Y.}~\bibnamefont {Li}},\ }\bibfield  {title} {\bibinfo {title} {An {Interior} {Trust} {Region} {Approach} for {Nonlinear} {Minimization} {Subject} to {Bounds}},\ }\href {https://doi.org/10.1137/0806023} {\bibfield  {journal} {\bibinfo  {journal} {SIAM Journal on Optimization}\ }\textbf {\bibinfo {volume} {6}},\ \bibinfo {pages} {418} (\bibinfo {year} {1996})}\BibitemShut {NoStop}%
\bibitem [{\citenamefont {Furthm{\"u}ller}\ \emph {et~al.}(1996)\citenamefont {Furthm{\"u}ller}, \citenamefont {Hafner},\ and\ \citenamefont {Kresse}}]{furthmuller1996dimer}%
  \BibitemOpen
  \bibfield  {author} {\bibinfo {author} {\bibfnamefont {J.}~\bibnamefont {Furthm{\"u}ller}}, \bibinfo {author} {\bibfnamefont {J.}~\bibnamefont {Hafner}},\ and\ \bibinfo {author} {\bibfnamefont {G.}~\bibnamefont {Kresse}},\ }\bibfield  {title} {\bibinfo {title} {Dimer reconstruction and electronic surface states on clean and hydrogenated diamond (100) surfaces},\ }\href {https://doi.org/10.1103/PhysRevB.53.7334} {\bibfield  {journal} {\bibinfo  {journal} {Phys. Rev. B}\ }\textbf {\bibinfo {volume} {53}},\ \bibinfo {pages} {7334} (\bibinfo {year} {1996})}\BibitemShut {NoStop}%
\bibitem [{\citenamefont {Kresse}\ and\ \citenamefont {Furthm{\"u}ller}(1996)}]{kresse1996efficiency}%
  \BibitemOpen
  \bibfield  {author} {\bibinfo {author} {\bibfnamefont {G.}~\bibnamefont {Kresse}}\ and\ \bibinfo {author} {\bibfnamefont {J.}~\bibnamefont {Furthm{\"u}ller}},\ }\bibfield  {title} {\bibinfo {title} {Efficiency of ab-initio total energy calculations for metals and semiconductors using a plane-wave basis set},\ }\href {https://doi.org/10.1016/0927-0256(96)00008-0} {\bibfield  {journal} {\bibinfo  {journal} {Comput. Mater. Sci.}\ }\textbf {\bibinfo {volume} {6}},\ \bibinfo {pages} {15} (\bibinfo {year} {1996})}\BibitemShut {NoStop}%
\bibitem [{\citenamefont {Perdew}\ \emph {et~al.}(1996)\citenamefont {Perdew}, \citenamefont {Burke},\ and\ \citenamefont {Ernzerhof}}]{perdew1996generalized}%
  \BibitemOpen
  \bibfield  {author} {\bibinfo {author} {\bibfnamefont {J.~P.}\ \bibnamefont {Perdew}}, \bibinfo {author} {\bibfnamefont {K.}~\bibnamefont {Burke}},\ and\ \bibinfo {author} {\bibfnamefont {M.}~\bibnamefont {Ernzerhof}},\ }\bibfield  {title} {\bibinfo {title} {Generalized gradient approximation made simple},\ }\href {https://doi.org/10.1103/PhysRevLett.77.3865} {\bibfield  {journal} {\bibinfo  {journal} {Phys. Rev. Lett.}\ }\textbf {\bibinfo {volume} {77}},\ \bibinfo {pages} {3865} (\bibinfo {year} {1996})}\BibitemShut {NoStop}%
\bibitem [{\citenamefont {Bl{\"o}chl}(1994)}]{blochl1994projector}%
  \BibitemOpen
  \bibfield  {author} {\bibinfo {author} {\bibfnamefont {P.~E.}\ \bibnamefont {Bl{\"o}chl}},\ }\bibfield  {title} {\bibinfo {title} {Projector augmented-wave method},\ }\href {https://doi.org/10.1103/PhysRevB.50.17953} {\bibfield  {journal} {\bibinfo  {journal} {Phys. Rev. B}\ }\textbf {\bibinfo {volume} {50}},\ \bibinfo {pages} {17953} (\bibinfo {year} {1994})}\BibitemShut {NoStop}%
\bibitem [{\citenamefont {Heyd}\ \emph {et~al.}(2003)\citenamefont {Heyd}, \citenamefont {Scuseria},\ and\ \citenamefont {Ernzerhof}}]{heyd2003hybrid}%
  \BibitemOpen
  \bibfield  {author} {\bibinfo {author} {\bibfnamefont {J.}~\bibnamefont {Heyd}}, \bibinfo {author} {\bibfnamefont {G.~E.}\ \bibnamefont {Scuseria}},\ and\ \bibinfo {author} {\bibfnamefont {M.}~\bibnamefont {Ernzerhof}},\ }\bibfield  {title} {\bibinfo {title} {Hybrid functionals based on a screened coulomb potential},\ }\href {https://doi.org/10.1063/1.1564060} {\bibfield  {journal} {\bibinfo  {journal} {J. Chem. Phys.}\ }\textbf {\bibinfo {volume} {118}},\ \bibinfo {pages} {8207} (\bibinfo {year} {2003})}\BibitemShut {NoStop}%
\end{thebibliography}%

\end{document}